\documentclass[twocolumn,amsmath,aps,fleqn]{revtex4}
 
\usepackage{amssymb}
\usepackage{amsmath}
\usepackage{epsfig}
\usepackage{subfigure}
\usepackage{mathrsfs}
\usepackage{longtable}
\usepackage{enumerate}
\usepackage{multirow}
\usepackage{threeparttable}
 
\newcommand{\be}{\begin{eqnarray}}
\newcommand{\ee}{\end{eqnarray}}
\newcommand{\Hu}{{\cal H}}
\newcommand{\FT}{{\cal F}_T}
\newcommand{\FTT}{{\tilde{\cal F}}_T}
\newcommand{\GT}{{\cal G}_T}
\newcommand{\GTT}{{\tilde{\cal G}}_T}
\newcommand{\rtg}{\sqrt{-g}}

 \allowdisplaybreaks[1]

\begin{document}

\title{The Parameterized Post-Friedmann Framework for Theories of Modified Gravity: Concepts, Formalism and Examples.}

\author{Tessa Baker}
\email{tessa.baker@astro.ox.ac.uk}
\affiliation{Astrophysics, University of Oxford, Denys Wilkinson Building, Keble Road, Oxford, OX1 3RH, UK}

\author{Pedro Ferreira}
\email{p.ferreira1@physics.ox.ac.uk}
\affiliation{Astrophysics, University of Oxford, Denys Wilkinson Building, Keble Road, Oxford, OX1 3RH, UK}

\author{Constantinos Skordis}
\email{skordis@nottingham.ac.uk}
\affiliation{School of Physics and Astronomy, University of Nottingham, University Park, Nottingham, NG7
2RD, UK}


\begin{abstract}
A unified framework for theories of modified gravity will be an essential tool for interpreting the forthcoming deluge of cosmological data. We present such a formalism, the Parameterized Post-Friedmann framework (PPF), which parameterizes the cosmological perturbation theory of a wide variety of modified gravity models. PPF is able to handle spin-0 degrees of freedom from new scalar, vector and tensor fields, meaning that it is not restricted to simple models based solely on cosmological scalar fields. A direct correspondence is maintained between the parameterization and the underlying space of theories, which allows us to build up a `dictionary' of modified gravity theories and their PPF correspondences. In this paper we describe the construction of the parameterization and demonstrate its use through a number of worked examples relevant to the current literature. We indicate how the formalism will be implemented numerically, so that the dictionary of modified gravity can be pitted against forthcoming observations. 
\end{abstract}

\maketitle

\section{Introduction}
\label{section:intro}

Einstein's theory of General Relativity has been required to defend its title as the true theory of gravitation since its birth. During most of the 20th century modifications to General Relativity (GR) were largely an abstract venture into the realms of mathematical possibility. Today the issue is a more pressing one that we have been forced to entertain by experimental results.

It is unfortunate, then, that our methods of testing GR have become \textit{less} efficient. Constraining modified theories on an individual basis is likely to be an infinite process, unless our ingenuity at constructing new theories wanes \cite{2012PhR...513....1C}. We need a fast way to test and rule out theories if we are to drive their population into decline. 

An analogous situation existed in the 1970s, when the question at hand was ``Is GR the correct description of gravity in the Solar System?" A compelling answer in the affirmative was provided by the Parameterized Post-Newtonian framework (PPN) \cite{ThorneWill1971,Will1971,Will2006}. In this formalism, competing gravitational theories were mapped onto a unified parameterization and constrained \textit{simultaneously} using data from laboratory tests, lunar laser ranging and early satellite experiments. 

We propose to revive this approach by constructing a parameterized framework that can be used to test the concept of cosmological modified gravity in a very general, model-independent way (the PPN framework itself cannot be applied on cosmological scales, though see \cite{Noh:2012ui} for related ideas). A parametric approach is nothing new; there has been a substantial body of work along these lines in recent years \cite{Hu:2007fw,Bertschinger:2008bb,Linder:2008uv, Skordis:2009bl, Pogosian:2010hi,Daniel:2010gt,Bean:2010kk,Baker2011, Dossett:2011vu,Thomas:2011tc,Zuntz2012,BattyePearson,Bloomfield1,Bloomfield2, Mueller_Bean_2012}. However, most approaches have considered modifications to the field equations of GR that are motivated by simplicity and their relevance  to a limited number of cases. Model-builders have moved beyond simple scalar-field theories, and there is need for a parameterization that can handle more sophisticated theories.
 
In this work we present a new formalism called the `Parameterized Post-Friedmann' framework (PPF), which systematically accounts for the limited number of ways in which the Einstein field equations can be modified at the linearized level. This means that it encapsulates a very wide variety of theories, without the use of approximations or numerical solutions. Table \ref{table:theories} gives a non-exhaustive list of theories that are covered by the PPF framework. We note that the name PPF has previously been employed to refer to a \textit{different} formalism, see footnote \footnote{The phrase `Parameterized Post-Friedmann' was first introduced in \cite{Tegmark:2002cy}, then later used in \cite{Hu:2007fw} to refer to a different formalism. We have chosen to recycle the name yet again here (with kind permission from W. Hu) because we believe an analogy with PPN is a concise and broadly accurate way of describing our formalism. However, we must stress that the analogy with PPN should not be taken to extremes - hopefully the differences between PPN and PPF are made clear in this paper. We advise the reader to take care which usage of the name PPF is being referred to in other works.} for clarification.
\begin{center}
\begin{table*}[t]
\begin{tabular}{| c | c | c |}
\hline
\bf{Category} & \bf{Theory} & References\\ \hline
\multirow{13}{*}{Horndeski Theories} & Scalar-Tensor theory& \multirow{2}{*}{\cite{BransDicke,2001PhRvD..63f3504E}} \\
&(incl. Brans-Dicke) & \\ \cline{2-3}
& $f(R)$ gravity & \cite{2010RvMP...82..451S, deFelice:2010aj}\\ \cline{2-3} 
& $f({\cal G})$ theories & \cite{Nojiri2005, Amendola_GB_2006, Koivisto2007}\\ \cline{2-3} 
& Covariant Galileons & \cite{Nicolis2009,Burrage2011,Gao_2011}\\ \cline{2-3}
& The Fab Four & \cite{Charmousis2012, FFderiv,Bruneton2012,Copeland2012} \\ \cline{2-3}
& K-inflation and K-essence & \cite{Armendariz1999,kessence}\\ \cline{2-3}
& Generalized G-inflation & \cite{Kobayashi2011, Deffayetetal_2011}\\ \cline{2-3}
& Kinetic Gravity Braiding & \cite{KGB1,KGB2}\\ \cline{2-3} 
& Quintessence (incl.&  \multirow{2}{*}{\cite{Caldwell_Dave_Steinhardt,Amendola2000,Steinhardt_2003,Pettorino2008}}\\ 
& universally coupled models)&\\ \cline{2-3}
& Effective dark fluid & \cite{Hu:1998ky} \\ \hline 
\multirow{2}{*}{Lorentz-Violating theories} & Einstein-Aether theory & \cite{Jacobson:2004ba,Jacobson_review,Zlosnik2007,Zlosnik2008}\\ \cline{2-3}
& Ho\u{r}ava-Lifschitz theory & \cite{Horava2009, Sotiriou_HL_review} \\ \hline
\multirow{3}{*}{$>2$ new degrees of freedom}& DGP (4D effective theory) & \cite{2000PhLB..485..208D,Fang:2008kc} \\ \cline{2-3}
& EBI gravity & \cite{Banados:2008bc,Banados:2009ec,Banados2010,AvelinoFerreira2012,Celia2012}\\ \cline{2-3}
& TeVeS & \cite{Bekenstein:2004fz,Skordis:2006ic,Skordis:2008be}\\ \hline  
\end{tabular}
\caption{A non-exhaustive list of theories that are suitable for PPF parameterization. We will not treat all of these explicitly in the present paper. ${\cal G}=R^2-4R_{\mu\nu}R^{\mu\nu}+R_{\mu\nu\rho\sigma}R^{\mu\nu\rho\sigma}$ is the Gauss-Bonnet term.} 
\label{table:theories}
\end{table*}
\end{center}
\vspace{-9mm}

 The key feature of our parameterization is that it maintains a direct correspondence between the parameters \footnote{We will see in the next section that in an expanding universe one is forced to use time-dependent functions as `parameters', rather than constant numbers.} of the formalism and `known' theories. In this paper we will refer to a `known' theory as an established model for which field equations can be written down analytically. These are usually derived directly from a covariant action, though knowledge of the action itself is \textit{not} required for PPF. A known model is represented by a point in the space of all possible theories (or a small region if the theory contains variable parameters). In contrast, we will use the description `unknown' for a point in theory-space for which we do not possess the corresponding action. An unknown theory is characterized purely in terms of its PPF parameters. PPF can be used to make statements about unknown regions of theory-space in addition to the testing of known theories. Such statements could be of use in guiding model-builders to the most relevant regions of theory-space.

Any parameterization has limits of applicability, and PPF is no exception. The (fairly mild) assumptions underlying our formalism are stated in Table \ref{table:assumptions}. PPN and PPF are highly complementary in their coverage of different accessible gravitational regimes. PPN is restricted to weak-field regimes on scales sufficiently small that linear perturbation theory about the Minkowski metric is an accurate description of the spacetime. Unlike PPN, PPF is valid for arbitrary background metrics (such as the FRW metric) provided that perturbations to the curvature scalar remain small. PPF also assumes the validity of linear perturbation theory,  so it is applicable to large length-scales on which matter perturbations have not yet crossed the nonlinear threshold (indicated by $\delta_M (k_{nl})\sim 1$); note that this boundary evolves with redshift. 

 Perturbative expansions like PPN and PPF cannot be used in the nonlinear, strong-field regime inhabited by compact objects. However, this regime can still be subjected to parameterized tests of gravity via electromagnetic observations \cite{Johannsen2010,Johannsen:2012tq} and the Parameterized Post-Einsteinian framework (PPE) for gravitational waveforms \cite{Yunes_PPE,Cornish_PPE}. Note that despite the similarity in nomenclature, PPE is somewhat different to PPN and PPF, being a parameterization of observables rather than theories themselves.

The purpose of this paper is to present the formalism that will be used for our future results \cite{Bakerinprep} and demonstrate its use through a number of worked examples. We would like to politely suggest three strategies for guiding busy readers to the most relevant sections: 
\begin{enumerate}[i)]
\item The casually-interested reader is recommended to assimilate the basic concepts and structure of the parameterization from \textsection\ref{subsection:formalism_basics} and \textsection\ref{subsection:formalism_frame}, and glance at Table \ref{table:theories} to see some example theories covered by this formalism. 

\item A reader with a particular interest in one of the example theories listed in Table \ref{table:theories} may wish to additionally read \textsection\ref{subsection:formalism_invariance}-\ref{subsection:formalism_vars} to understand how the mapping into the PPF format is performed, and the most relevant example(s) of \textsection\ref{section:examples}.

\item A reader concerned with the concept of parameterized modified gravity in and of itself may also find \textsection\ref{subsection:formalism_constraints} and \textsection\ref{section:PPF_coeffs} useful for explaining how the approach presented here can be concretely implemented (for example, in numerical codes). \textsection\ref{section:PPF_coeffs} also discusses the connection of PPF to other parameterizations in the present literature.
\end{enumerate}
Our conclusions are summarized in \textsection\ref{section:conclusions}. 

We will use the notation $\kappa=M_P^{-2}=8\pi G$ and set $c=1$ unless stated otherwise. Our convention for the metric signature is \mbox{$(-,+,+,+)$}. Dots will be used to indicate differentiation with respect to conformal time and hatted variables indicate gauge-invariant combinations, which are formed by adding appropriate metric fluctuations to a perturbed quantity (see \textsection\ref{subsection:formalism_vars}).  Note that this means \mbox{$\hat\chi\neq\chi$}. 

\begin{center}
\begin{table*}[t]
\begin{tabular}{| l |}
\hline
\multicolumn{1}{| c |}{ \bf{Assumptions and Restrictions of the PPF Parameterization}}  \\ \hline 
All field equations are second-order or lower in time derivatives  (but $f(R)$ gravity is still treatable -- see text). \\ \hline
There exists an FRW solution for the background cosmology.$^{\dag}$\\ \hline 
 There exists a frame in which matter components obey their ordinary conservation equations, that is, $\nabla_\mu T^\mu_\nu=0$.$^{\ddag}$ \\ \hline
The field equations of a gravitational theory are gauge form-invariant (see \textsection\ref{subsection:formalism_invariance}).\\ \hline
Nonlinear perturbations are negligible at the lengthscales under consideration. \\ \hline
If $N$ non-GR fields are present and $N>2$, then $N-2$ relations between the new fields must be specified. If $N<2$ (the \\majority of cases) then no additional relations are required.  \\ \hline
\end{tabular}
\caption{ {A summary of the assumptions underlying the PPF formalism.}
\newline
$^{\dag}$ {\footnotesize This potentially poses a problem for Massive Gravity, in which exact, flat FRW solutions do not exist \cite{DAmico:2011ul}. However, it may still be possible to map the slightly perturbed \textit{background} solutions of \cite{Volkov2012} onto our parameterization.}\newline
$^{\ddag}$ {\footnotesize Models that posit a universal coupling between a quintessence field and matter components are treatable, see \textsection\ref{subsubsection:formalism_vars_matter}. Models which implement non-universal couplings are not.} }
\label{table:assumptions}
\end{table*}
\end{center}

\section{The PPF Formalism}
\label{section:formalism}

\subsection{Basic Principles}   
\label{subsection:formalism_basics}

As stated in the introduction, the PPF framework systematically accounts for allowable extensions to the Einstein field equations, whilst remaining agnostic about their precise form. A particularly important extension is to permit the existence of new scalar degrees of freedom (hereafter d.o.f.) that are not present in GR \footnote{This is the most common method that mainstream modified gravity theories use to evade Lovelock's theorem. Lovelock's theorem states that the only gravitational theory with second-order equations of motion and no new degrees of freedom (beyond those of GR) that can be derived from a covariant action in four dimensions is that of GR with a cosmological constant term. \newline
\indent However, some theories instead evade Lovelock's theorem by invoking more than four dimensions, nonlocality, or higher-derivative field equations.}. We emphasize that scalar d.o.f. are not synonymous with scalar fields, but can also arise from the spin-0 perturbations of vectors and tensors.
In addition to these possible new d.o.f., there could also be modifications involving the d.o.f. already present in GR. Therefore we need to allow for perturbations of the spacetime metric to appear in the linearized field equations in a non-standard way. 

With these concepts in mind, the construction of the PPF parameterization proceeds as follows: first, we add to the linearized gravitational field equations of GR new terms containing perturbations of all the scalar degrees of freedom that are present. Schematically:
\begin{eqnarray}
\label{schematic_U}
\delta G_{\nu}^{\mu}&=&\kappa \, \delta T_{\nu}^{\mu}+\delta U_{\nu}^{\mu}\\
\mathrm{where}\quad \delta U_{\mu\nu}&=& \delta U_{\mu\nu}^{\mathrm{metric}}(\delta g_{\rho\sigma})+ \delta U_{\mu\nu}^{\mathrm{d.o.f.}}(\delta\phi, V, \upsilon...) \nonumber\\
&+&\delta U_{\mu\nu}^{\mathrm{matter}}(\delta_M, \theta_M...)\nonumber
\end{eqnarray}
In the expression above the tensor of modifications $\delta U_{\mu\nu}$ has been decomposed into parts containing scalar perturbations of the metric, scalar perturbations of new d.o.f., and matter perturbations. We will see in \textsection\ref{subsubsection:formalism_vars_matter} that the matter term can be eliminated in favour of additional contributions to the first two terms of eq.(\ref{schematic_U}). 
Possible contributions to $\delta U_{\mu\nu}^{\mathrm{d.o.f.}}$ from new vector and tensor fields are indicated - schematically for the present - by $\delta A_{i}=(\vec{\nabla}_i V)/a$ and $2\upsilon \gamma_{ij}=-\delta \tilde{g}_{ij}$ respectively, where $\gamma_{ij}$ is a spatial matrix, $A_\mu$ is a new vector field and $\upsilon$ is a spin-0 perturbation of a new tensor field $\tilde{g}_{\rho\sigma}$. It is necessary to specify the number of new d.o.f. \textit{a priori}, but one can remain completely general as to their physical origin. 

Likewise we also need to choose the derivative order of the parameterization, that is, the highest number of time derivatives that appear in the field equations. We will choose this to be two, as Ostrogradski's theorem \cite{Ostrogradski} generically leads to the existence of instabilities in higher-derivative theories. Under special circumstances these instabilities can be avoided \cite{Woodard:2006cp}, such as occurs in the popular class of $f(R)$ gravity theories. However, any $f(R)$ theory can be mapped onto a second-order scalar-tensor theory via a Legendre transformation \cite{2010RvMP...82..451S}, so our parameterization is still applicable to $f(R)$ gravity in this form. A reminder of this transformation is given in Appendix \ref{appendix:fR_transf}.

The coefficient of each perturbation appearing in the PPF parameterization can only be a function of zeroth order `background' quantities; for an isotropic and homogeneous universe this means that they are only functions of time and scale (through Fourier wavenumber $k$). The scale-dependence of these coefficient functions is not completely arbitrary -- it is related to the derivative order of theory (see \textsection\ref{subsection:scale_dep}). We will find that for many known theories it has a simple polynomial form.

Having accounted for all terms permitted in a second-order theory, one then imposes \textit{gauge form-invariance} on the parameterization (to be explained momentarily in \textsection\ref{subsection:formalism_invariance}). 

One can imagine that the parameterization dictated by the above principles could potentially contain large numbers of free functions. Fortunately these are not all independent; for most cases there exist constraint equations that reduce their number considerably. These are discussed in \textsection\ref{subsection:formalism_constraints} below. 

Finally we highlight that as the PPF framework is based on the linearized field equations, it is applicable to scale regimes that are well above those affected by screening mechanisms. Effects such as the chameleon, symmetron or Vainshstein mechanisms \cite{2004PhRvL..93q1104K,2004PhRvD..69d4026K,2010arXiv1011.5909K,2010PhRvL.104w1301H,2011PhRvD..84j3521H,Brax_symmetron,Vainshtein1972} are not present in the parameterization.

\subsection{Gauge Form-Invariance}
\label{subsection:formalism_invariance}

Any equations of motion that can be derived from an action principle must possess the property of form-invariance, as we explain here in terms classical mechanics.

 A canonical transformation is defined to be one that takes us from a set of coordinates and their conjugate momenta $\{q_i,p_i\}$, to a second set $\{Q_i,P_i\}$ in such a way that the structure of Hamilton's equations for any dynamical system is preserved. Hamilton's equations are said to be form-invariant under canonical transformations. Here, form-invariance means that the equations of motion in the new coordinate system are simply those obtained by replacing $\{q_i,p_i\}$ with  $\{Q_i,P_i\}$ in the original system. This symmetry is not be confused with general covariance which, conceptually, enforces that the physics of a situation remains independent of coordinate choice.  General covariance allows one to choose a convenient set of coordinates such that some perturbation variables are set to zero, which will clearly cause the structure of the field equations to appear \textit{different} after the transformation \cite{Malik:2012tb}. 
 
 One can show \cite{HandFinch} that the form-invariance of Hamilton's equations carries over to their equivalent Lagrangian formulation. A gauge transformation \mbox{$x^\mu\rightarrow x^\mu+\xi^{\mu}$} is an example of a canonical transformation. Therefore, if a set of gravitational field equations are derivable as the Euler-Lagrange equations of an action then they must be form-invariant under gauge transformations of this kind: they are \textit{gauge form-invariant}.

Gauge form-invariance will be realized in our examples because any new terms containing the generating vector $\xi^\mu$ that arise from a transformation will cancel each other by virtue of the background-level field equations, leaving the original expression unchanged. See \textsection\ref{subsection:formalism_frame} for a demonstration of this. This property will not be `visible' if one has already fixed a gauge.

To reiterate more concisely: before we choose a gauge, the linearized field equations are gauge form-invariant; in a specific gauge they are not. 
Hence enforcing gauge form-invariance in our parameterization ensures that it does not implicitly correspond to any particular gauge choice. We implement this explicitly by combining as many terms as possible into gauge-invariant variables, and adding a gauge form-invariance-fixing term to guarantee that any remaining gauge-variant pieces cancel via the background field equations. Note that this does not represent a sleight-of-hand in any sense, as we know that the true underlying field equations must be gauge form-invariant.

\subsection{Background-Level Parameterization}
\label{subsection:formalism_background}

In eq.(\ref{schematic_U}) we introduced a tensor of modifications to the field equations of GR. Let the zeroth-order components of this tensor be \mbox{$U_{0}^0=-X,\,U^i_j=Y\delta^i_j$}, such that the modified Friedmann and Raychaudhuri equations are:
\be
3\left(\Hu^2+K\right)&=&\kappa a^2\rho_M+a^2X \label{Friedmann}\\
2\left(\Hu^2-\dot\Hu+K\right)&=&\kappa a^2 \rho_M(1+\omega_M)+a^2(X+Y) \label{Raych}
\ee
where $\Hu$ is the conformal Hubble factor and dots denote derivatives with respect to conformal time. Hereafter we specialize to flat cosmologies, setting $K=0$.  $\rho_M$ denotes a sum over standard matter components, including cold dark matter, with the effective total equation of state $\omega_M$. Modifications to GR and exotic fluids/dark energy are formally indistinguishable at the unperturbed level, so $X$ and $Y$ can be regarded as an effective energy density and pressure respectively.

\subsection{Perturbation Variables}
\label{subsection:formalism_vars}
\subsubsection{Metric Variables}
We write the perturbed line element as (showing scalar perturbations only):
\be
\label{line_element}
ds^2 &=& a(\eta)^2\Big[-(1-2 \Xi)d\eta^2-2(\vec{\nabla}_i\epsilon)d\eta\,dx^i\nonumber\\
&+&\left(1+\frac{1}{3}\beta\right)\gamma_{ij}+\left(D_{ij}\nu\right) \,dx^i dx^j \Big]
\ee
where $\gamma_{ij}$ is a flat spatial metric and $D_{ij}$ is a derivative operator that projects out the longitudinal, traceless, spatial part of the perturbation:
\begin{align}
D_{ij}&=\vec{\nabla}_i\vec{\nabla}_j-1/3 \gamma_{ij}\vec{\nabla}^2
\end{align}
Bardeen introduced two gauge-invariant combinations of the metric perturbations \cite{Bardeen:1980td}: 
\begin{align}
\label{Phi_def}
\hat\Phi&=-\frac{1}{6}(\beta-\nabla^2\nu)+\frac{1}{2}\Hu(\dot\nu+2\epsilon)  \\
\hat\Psi&=-\Xi-\frac{1}{2}(\ddot\nu+2\dot\epsilon)-\frac{1}{2}\Hu (\dot\nu+2\epsilon) \label{Psi_def}
\end{align}
These reduce to the familiar potentials $\Phi$ and $\Psi$ of the conformal Newtonian gauge upon setting \mbox{$\epsilon=\nu=0$} -- though beware different sign conventions. For our purposes it will be more convenient to use gauge-invariant variables of the same derivative order (note that $\hat\Psi$ is second-order in time derivatives, whilst $\hat\Phi$ is first-order). For this reason we define the following combination, in which the second-order terms cancel:
\begin{equation}
\hat\Gamma=\frac{1}{k}\left(\dot{\hat\Phi}+\Hu\hat\Psi\right)
\label{Gamma_def}
\end{equation}
$\hat\Phi$ and $\hat\Gamma$ will be the basic building blocks of the metric sector of our parameterization. 

\subsubsection{New Degrees of Freedom}
\label{subsubsection:formalism_dof}
We also need to introduce a gauge form-invariant way of parameterizing the new scalar degrees of freedom. First consider a gauge transformation \mbox{$x^\mu\rightarrow x^\mu+\xi^\mu$} generated by the vector \mbox{$\xi^\mu=1/a\left(\xi^0,\,\nabla^i \psi\right)$}. The most general way that a dimensionless scalar perturbation $\chi$ can transform under this co-ordinate shift is:
\begin{equation}
\label{chi_transf}
\chi\rightarrow\chi+\frac{1}{a}\left(G_1\xi^0+G_2\dot\xi^0+G_3\psi+G_4\dot\psi\right)
\end{equation}
where $G_1-G_4$ are functions that are fixed in a known theory, and we have suppressed their arguments. For example, the perturbation of a standard scalar field transforms as \mbox{$\delta\phi\rightarrow \delta\phi+\dot\phi\, \xi^0/a$}, so in this instance $G_1=\dot\phi$ and $G_2=G_3=G_4=0$ (we have implicitly normalized $\phi$ by a mass scale to reduce clutter whilst keeping $\chi$ dimensionless). Table \ref{table:dof_transfs} gives the transformation properties of some other common types of scalar degrees of freedom.
\begin{center}
\begin{table*}
\begin{tabular}{| c | c |  c  c  c  c |}\hline
\bf{Perturbation Type} & \,\bf{Symbol}\, & $\bf{G_1}$ & $\bf{G_2}$ & $\bf{G_3}$ & $\bf{G_4}$ \\ \hline
\multirow{4}{*}{Metric } & $\Xi$ & 0 & -1 & 0 & 0 \\
 & $\epsilon$ & 1 & 0 & $\Hu$ & -1 \\ 
 & $\beta$ & 6$\Hu$ & 0 & -2$k^2$ & 0 \\ 
 & $\nu$ & 0 & 0 & 2 & 0 \\ \hline
 {Fractional energy density} & $\delta$ & -3$\Hu (1+\omega)$ & 0 & 0 & 0 \\
 {Velocity potential} & $\theta$ & 1 & 0 & 0 & 0 \\   
 {Fractional pressure} & $\Pi$ & $-3\Hu(1+\omega)c_a^2$$^{\dag}$ & 0 & 0 & 0 \\
 {Anisotropic stress} & $\Sigma$ & 0 & 0 & 0 & 0 \\ \hline
 Scalar field & $\delta\phi$ & $\dot\phi$ & 0 & 0 & 0 \\ \hline
 Timelike vector: spatial component$^{\ddag}$  & ${V}$ & 0 & 0 & $\Hu$ & -1 \\ \hline
\end{tabular}
\caption{The behaviour of perturbations encountered in this paper under the gauge transformation \mbox{$x^\mu\rightarrow x^\mu+\xi^\mu$}, where \mbox{$\xi^{\mu}=1/a (\xi^0,\vec{\nabla}^i\psi)$}.\newline
$^{\dag}${\footnotesize $c_a^2=\omega-\frac{\dot\omega}{3\Hu (1+\omega)}$ is the adiabatic sound speed. }\newline
$^{\ddag}${\footnotesize Such that the perturbations to the spatial components of the vector are written $\delta A^i=(\vec{\nabla}^i V)/a.$}} 
\label{table:dof_transfs}   
\end{table*}
\end{center}

We want to find a combination of the new d.o.f. and metric perturbations such that the gauge transformation properties of each term cancel to zero overall. An infinite number of such possibilities exists, but we will impose the restriction that the resulting gauge-invariant combination contains no time derivatives. Then our gauge invariant variable must have the form:
\begin{equation}
\label{chi_hat_omega}
\hat\chi=\chi+\omega_1\Xi+\omega_2\beta+\omega_3\epsilon+\omega_4\nu
\end{equation}
By using the transformation properties given in eq.(\ref{chi_transf}) and Table \ref{table:dof_transfs}, and requiring that $\hat\chi$ has no net transformation, one finds:
\begin{align}
\label{omegas}
\omega_1&=G_2 &  \omega_2&=-\frac{1}{6\Hu}(G_1+G_4)\nonumber\\
\omega_3&=G_4 & \omega_4&=-\frac{1}{2}(G_3+\Hu G_4)-\frac{k^2}{6\Hu}(G_1+G_4)  
\end{align}
In the case of a standard scalar field one obtains \mbox{$\hat\chi=\delta\phi-\frac{\dot\phi}{6\Hu}(\beta+k^2\nu)$}. Eqs.(\ref{chi_hat_omega}) and (\ref{omegas}) should be thought of a `template' gauge-invariant variable that can be adapted to play the role of the new d.o.f. in a given theory. Note that if one uses a perturbation of the spacetime metric as an input to this prescription the result vanishes, since $\hat\chi$ represents \textit{new} d.o.f. only.

\subsubsection{Matter Perturbations}
\label{subsubsection:formalism_vars_matter}
A coupling between new d.o.f. and matter can give rise to an effective evolving and/or scale-dependent gravitational constant, e.g. in a typical scalar-tensor theory  $G_{\mathrm{eff}}=G/\phi$. This means that when we write the modified field equations in the form of eq.(\ref{schematic_U}) $\delta U^\mu_\nu$ will contain matter variables. For example, consider a case where the modified field equations have the following form:
\begin{equation}
\label{Q_FEs}
\delta G_{\mu\nu}=\frac{\kappa G}{f(\phi, A_\rho, \tilde{g}_{\rho\sigma}...)}\delta T_{\mu\nu}+\left[\delta U^{\mathrm{dof}}_{\mu\nu}+\delta U_{\mu\nu}^{\mathrm{metric}}\right]
\end{equation}
where once more we have indicated that the modifications could arise from new scalar, vector or tensor fields, and some arguments have been suppressed.
The renormalized matter terms are taken into $\delta U_{\mu\nu}$:
\begin{equation}
\delta U_{\mu\nu}=\left(\frac{1}{f(\phi, A_\mu, \tilde{g}_{\mu\nu}...)}-1\right)\kappa \delta T_{\mu\nu}+\left[\delta U^{\mathrm{dof}}_{\mu\nu}+\delta U_{\mu\nu}^{\mathrm{metric}}\right]
\end{equation}
$\delta T_{\mu\nu}$ can then be eliminated from the expression above using eq.(\ref{schematic_U}). Rearranging, we then have $\delta U_{\mu\nu}$ expressed entirely in terms of metric variables and new d.o.f.:
\begin{equation}
\label{shuffled_U}
\delta U_{\mu\nu}=\left[1-f\right] \delta G_{\mu\nu}+ \,f \left[\delta U^{\mathrm{dof}}_{\mu\nu}+\delta U_{\mu\nu}^{\mathrm{metric}}\right]
\end{equation}

A theory that posits a universal coupling between a new d.o.f. and matter species can be transformed to a frame where the coupling is absent. These are not necessarily the standard Einstein and Jordan frames, as the field equations do not have to match GR in either frame. In the uncoupled frame the the modification tensor $U_{\mu\nu}$ will be independently conserved, and the PPF parameterization can proceed as normal (we explain why the relation \mbox{$\nabla_\mu U^\mu_\nu=0$} is necessary in \textsection\ref{subsection:formalism_constraints}).

In the case of non-universal couplings \cite{Amendola2000, Pettorino2008} it may not be possible to find a frame in which the relations \mbox{$\nabla_\mu U^\mu_\nu=0$} and \mbox{$\nabla_\mu T^\mu_\nu=0$} hold individually. Our parameterization cannot be applied to these cases in its present formulation.

\subsection{Perturbation Framework}
\label{subsection:formalism_frame}

The components of the perturbed Einstein tensor are:
\begin{align}
-a^2\delta G_0^0 &=E_{\Delta} = 2\nabla^2\hat\Phi-6{\cal H}k\hat\Gamma-3\Hu (\Hu^2-\dot\Hu)(\dot\nu+2\epsilon)  \nonumber \\
-a^2\delta G^0_i &=\nabla_i E_{\Theta}  = 2\,k\hat\Gamma+ (\Hu^2-\dot\Hu)(\dot\nu+2\epsilon) \nonumber\\
a^2\delta G^i_i &=E_P = 6k\dot{\hat\Gamma}+12{\cal H}k\hat\Gamma-2\vec{\nabla}^2(\hat\Phi-\hat\Psi)\nonumber \\*
&   -6 (\Hu^2-\dot\Hu)\hat\Psi+3\left(\Hu^3+\dot\Hu \Hu-\ddot\Hu\right)(\dot\nu+2\epsilon)\nonumber\\
a^2\delta \tilde{G}^i_j &=D^i_jE_{\Sigma}  = D^i_j(\hat\Phi-\hat\Psi)\label{es}
\end{align}
where 
\begin{align}
\delta \tilde{G}^i_j=\delta G^i_j-\frac{\delta^i_j}{3}\delta G^k_k  
\end{align}
Recall that $D^i_j$ projects out the longitudinal, traceless, spatial part of $\delta G^{\mu}_{\nu}$; we have defined the quantities $E_i$ in order to refer to the left-hand side of the Einstein equations with ease. We will also use analogous symbols to refer to the components of the $\delta U^\mu_\nu$ tensor, i.e.
\begin{align}
U_{\Delta}&=-a^2\delta U^0_0  & \vec{\nabla}_i U_{\Theta}&=-a^2 \delta U^0_i\nonumber\\
 U_P&=a^2\delta U^i_i & D^i_jU_{\Sigma}&=a^2(U^i_j-\frac{1}{3}\delta U^k_k\delta^i_j)
\label{U_components_def}
\end{align}

Having eliminated any matter perturbations from $\delta U^\mu_\nu$ via the method described in the previous subsection, we can expand $\delta U^\mu_\nu$ in terms of the gauge-invariant variables of eqs.(\ref{Phi_def}), (\ref{Gamma_def}) and (\ref{chi_hat_omega}). We write down all terms permitted in a second-order theory. We will see in \textsection \ref{subsection:formalism_constraints} that the equation(s) of motion (hereafter e.o.m.) for the new d.o.f. are obtained from the Bianchi identities, and require differentiation of $\delta U^0_0$ and $\delta U^0_i$; hence these components can only contain first-order time derivatives if the e.o.m.s are to remain at second order.

For simplicity we will consider the case where only one new scalar d.o.f. is present, but the extension to multiple d.o.f. is straightforward. The PPF parameterization of the modified field equations is then as follows:
\begin{widetext}
 \begin{align}
E_{\Delta}&=\kappa a^2 G\,\rho_M\delta_M+A_0 k^2\hat\Phi+F_0k^2\hat\Gamma+\alpha_0k^2\hat\chi+\alpha_1k\dot{\hat\chi}+k^3 M_{\Delta}(\dot\nu+2\epsilon)\label{FE1}\\ 
E_{\Theta}&=\kappa a^2 G\,\rho_M (1+\omega_M)\theta_M+B_0 k\hat\Phi+I_0k\hat\Gamma+\beta_0 k\hat\chi+\beta_1\dot{\hat\chi}+k^2M_{\Theta}(\dot\nu+2\epsilon)\label{FE2}\\
E_P&=3\,\kappa a^2 G\,\rho_M\Pi_M+C_0 k^2\hat\Phi+C_1 k\dot{\hat\Phi}+J_0k^2\hat\Gamma+J_1 k\dot{\hat\Gamma}+\gamma_0 k^2\hat\chi+\gamma_1 k \dot{\hat\chi}+\gamma_2 \ddot{\hat\chi}+k^3M_P (\dot\nu+2\epsilon)\label{FE3}\\
E_{\Sigma}&=\kappa a^2 G\,\rho_M (1+\omega_M)\Sigma_M+ D_0\hat\Phi+\frac{D_1}{k} \dot{\hat\Phi}+K_0\hat\Gamma+\frac{K_1}{k}\dot{\hat\Gamma}+\epsilon_0\hat\chi+\frac{\epsilon_1}{k}\dot{\hat\chi}+\frac{\epsilon_2}{k^2} \ddot{\hat\chi}  \label{FE4}
\end{align}
\end{widetext}
The coefficients $A_0 -\epsilon_2$ appearing in these expressions are \textit{not} constants - they are functions of the cosmological background, i.e. functions of time and scale. These dependencies have been suppressed above for the sake of clarity. The factors of $k$ (Fourier wavenumber) accompanying each term are such that $A_0-\epsilon_2$ are all dimensionless. A particular known theory will specify exact functional forms for these coefficients; they can be considered as the PPF equivalent of the ten PPN parameters. They are the `slots' one maps a theory of modified gravity onto.

$M_{\Delta}$, $M_{\Theta}$ and $M_P$ are the gauge form-invariance-fixing terms described in \textsection\ref{subsection:formalism_invariance}, and are similarly functions of background variables. However, the $M_i$ differ from the coefficients  $A_0 -\epsilon_2$ in that the former are fixed by the zeroth-order field equations, whilst the latter are determined by the linearly perturbed field equations. 

Let us demonstrate this explicitly for eq.(\ref{FE1}). Under gauge transformations of the form:
\begin{align}
x^\mu&\rightarrow x^\mu+\xi^\mu & \xi^\mu&=1/a\left(\xi^0,\,\nabla^i \psi\right)
\end{align}
 eq.(\ref{FE1}) becomes:
\begin{eqnarray}
E_{\Delta}^{\prime}-6\Hu(\Hu^2-\dot\Hu)\frac{\xi^0}{a}&=&\kappa a^2\rho_M\left[\delta_M^{\prime}-3\Hu(1+\omega_M)\frac{\xi^0}{a}\right]\nonumber\\
&+&k^3M_{\Delta}\left[(\dot\nu^{\prime}+2\epsilon^{\prime})+\frac{2\xi^0}{a}\right]
\end{eqnarray}
 where primed variables indicate those belonging to the transformed coordinate system. In order for the form of eq.(\ref{FE1}) to be exactly preserved by the transformation the terms containing $\xi^0$ must cancel, which determines $M_{\Delta}$:
 \begin{eqnarray}
 M_{\Delta}&=&-\frac{3\Hu}{2k^3}\left[2(\Hu^2-\dot\Hu)-\kappa a^2\rho_M (1+\omega_M)\right] \nonumber\\
 &=&-\frac{3\Hu}{2k^3}a^2\left(X+Y\right)
 \end{eqnarray}
 where eq.(\ref{Raych}) has been used in reaching the second equality. Analogously one finds:
 \begin{align}
 M_{\Theta}&=\frac{1}{2k^2}a^2(X+Y) & M_P &=\frac{3}{2k^3}a^2\dot{Y}
 \end{align}
 No gauge form-invariance-fixing term is needed in eq.(\ref{FE4}) because all the terms there are individually gauge invariant, including the matter shear perturbation $\Sigma_M$.
 
 Note that the $M_i$ are unlikely to be useful discriminators between theories because their evolution with redshift should be broadly similar in theories that reproduce a $\Lambda$CDM-like background. It is in the coefficients $A_0 - \epsilon_2$ that the differences between theories will be manifested.

\subsection{Constraint Equations}
\label{subsection:formalism_constraints}

The parameterization laid out in eqs.(\ref{FE1})-(\ref{FE4}) contains twenty-two coefficient functions, $A_0 - \epsilon_2$. We will see in the worked examples of  \textsection\ref{section:examples} that these coefficients are often very simply related and not independent. However, the purpose of PPF is to obtain non-model-specific constraints from the data - does this mean that we must run a Markov Chain Monte Carlo analysis with twenty-two free functions? It would be justifiable to object that current and near-future data may not possess sufficient constraining power for such a task.

Fortunately this is not the case for the analyses most relevant to current research. Imposing some restrictions on the types of theories being considered allows one to derive constraint relations between the PPF coefficients, thereby immediately eliminating some freedom from the parameterization. We will present a set of seven constraints relations here; it is very likely that others exist, which we will leave for future investigation \cite{Bakerinprep}.

This set of seven constraint relations stems from the divergenceless nature of the Einstein tensor, \mbox{$\nabla_\mu G^{\mu}_\nu=0$} (a straightforward consequence of the Bianchi identities). We will assume that ordinary matter obeys its standard conservation law, \mbox{$\nabla_\mu T^{\mu}_\nu=0$}. Therefore the U-tensor must also be divergenceless, \mbox{$\nabla_\mu U^{\mu}_\nu=0$}; the result \mbox{$\delta\left(\nabla_\mu U^{\mu}_\nu\right)=0$} follows at the perturbative level. Note that this statement is not valid for models that involve a non-universal coupling between a quintessence field and matter species, see \textsection\ref{subsubsection:formalism_vars_matter}. 

In an isotropic spacetime the expression \mbox{$\delta\left(\nabla_\mu U^{\mu}_\nu\right)=0$} has two independent components (for $\nu=0,\,i$) and so yields two second-order equations that specify how the evolution of $\hat\chi$ is tied to the metric potentials. However, this situation is potentially problematic - how can we guarantee that the solutions of these two equations agree? In a known theory with a single new field there will be \textit{one} equation of motion for $\hat\chi$. How do we reconcile this fact with the \textit{two} evolution equations coming from  \mbox{$\delta\left(\nabla_\mu U^{\mu}_\nu\right)=0$} ?

There are three possible ways that the problem could be resolved:
\begin{enumerate}
\item The true e.o.m. is given by the $\nu=0$ component of the Bianchi identity. The $\nu=i$ component reduces to a triviality because all of the coefficients of $\hat\chi,\,\hat\Phi$ and $\hat\Gamma$ in it vanish identically.
\item Instead the converse is true: the $\nu=i$ component becomes the e.o.m., and the $\nu=0$ component reduces to a triviality.
\item The true e.o.m. corresponds to a combination of the $\nu=0,\,i$ components. In this case one set of solutions for \{$\hat\chi,\,\hat\Phi,\,\hat\Gamma$\} would necessarily have to solve both equations.
\end{enumerate}
We will categorize theories as `type 1', `type 2' etc. according to which of the three possibilities above occurs. All of the single-field theories we tackle in this paper are type 1 theories, except for the special case of \textsection\ref{subsection:examples_fluid} as discussed below. 
This is what one might intuitively expect - we often think of the e.o.m. of a field as an expression of conservation of its energy, and it is the $\nu=0$ component of \mbox{$\nabla_\mu T^{\mu}_\nu=0$} that reduces to the energy conservation statement of a classical fluid in Minkowski space.

One can also construct an argument as to why \mbox{type 1} theories are the most natural occurrence by deriving the conservation law for $U_{\mu\nu}$ at the level of the action. In this derivation the statement $\nabla_{\mu}U^\mu_0=0$ is obtained by considering translation along a timelike gauge vector, which corresponds to evolving the new fields along a worldline. The statement $\nabla_{\mu}U^\mu_i=0$ arises from translation along the spacelike Killing vectors. As this corresponds to a symmetry of the FRW spacetime, it should not lead us to dynamical equations. See Appendix \ref{appendix:U_proof} for more details.

For a type 1 theory, then, the e.o.m. in terms of the parameterization of eqs.(\ref{FE1}-\ref{FE4}) is:
\begin{align}
\label{Bianchi_1}
&\left[\alpha_1+\Hu_k\gamma_2\right]\ddot{\hat\chi}+\left[\alpha_0+\beta_1+\frac{\dot{\alpha}_1}{k}+\Hu_k (\alpha_1+\gamma_1)\right]\,k\,\dot{\hat\chi} \nonumber\\
+&\left[\frac{\dot{\alpha}_0}{k}+\Hu_k (\alpha_0+\gamma_0)+\beta_0\right]\,k^2\,\hat\chi \nonumber\\
+ &\left[A_0+\Hu_k C_1-3\frac{a^2}{k^2}(X+Y)\right] k\,\dot{\hat\Phi} \nonumber \\
+&\left[\frac{\dot{A}_0}{k}+\Hu_k (A_0+C_0)+B_0\right]\,k^2\,\hat\Phi+\left[F_0+\Hu_k J_1\right]\,k\,\dot{\hat\Gamma}\nonumber\\
+&\left[\frac{\dot{F}_0}{k}+\Hu_k (F_0+J_0)+I_0\right]\,k^2\,\hat\Gamma=0
\end{align} 
where $\Hu_k=\Hu/k$. Meanwhile the seven coefficients of $\{\hat\Phi,\,\dot{\hat\Phi},\,\hat\Gamma,\,\dot{\hat\Gamma},\,\hat\chi,\,\dot{\hat\chi},\,\ddot{\hat\chi}\}$ in the second Bianchi component \mbox{$(\nu=i)$} must all vanish identically, leading to the following set of constraint equations:
\begin{align}
\label{Bianchi_2_constraints_start}
&\beta_1-\frac{\gamma_2}{3}+\frac{2}{3}\epsilon_2=0\\
&\beta_0+\frac{1}{k}(\dot\beta_1+2\Hu\beta_1)-\frac{1}{3}(\gamma_1-2\epsilon_1)=0 \\
&\frac{\dot{\beta}_0}{k}+2\Hu_k\beta_0-\frac{1}{3}(\gamma_0-2\epsilon_0)=0 \\
&B_0-\frac{1}{3}\left(C_1-2 D_1\right)+\frac{1}{\Hu k}a^2 (X+Y)=0\\
&\frac{\dot{B}_0}{k}+2\Hu_k B_0-\frac{1}{3}(C_0-2 D_0)=0 \\
&I_0-\frac{1}{3}\left(J_1-2K_1\right)=0\\
&\frac{\dot{I}_0}{k}+2\Hu_k I_0-\frac{1}{3}(J_0-2 K_0)-\frac{1}{\Hu k}a^2 (X+Y)=0
\label{Bianchi_2_constraints_end}
\end{align}
Thus if one is prepared to focus on type 1 theories - which covers a large part of current investigations into modified gravity - then seven free functions are immediately removed from the PPF parameterization. 

Theories with more than one new d.o.f. cannot be classified as type 1/2/3. For a theory with precisely two new d.o.f. it is necessary to use \textit{both} components of the conservation law for $U_{\mu\nu}$ to obtain the required two evolution equations. However, if one of the evolution equations can be suitably inverted it may be possible to eliminate one of the two extra fields; this will modify the PPF coefficients of the original system. The new system can then be classified as type 1/2/3 if either/neither Bianchi component reduces to a triviality when the \textit{new} PPF coefficients are used. An example of this is given in \textsection\ref{subsection:examples_fluid}.

For theories with more than two new d.o.f. - for example DGP (\textsection\ref{subsection:examples_dgp}) and Eddington-Born-Infeld gravity  (\textsection\ref{subsection:examples_ebi}) - extra relations between the new fields must be provided to close the system of equations. This will prevent us from being able to constrain regions of \textit{unknown} theory space with more than two new fields. Fortunately, based on the relative scarcity of such theories in the literature - and simplicity arguments - this type of theory is not of primary interest at present.

\section{Worked Examples}
\label{section:examples}

In this section we demonstrate how a variety of commonly-discussed theories of modified gravity can be mapped onto our parameterization. We will not discuss the motivation or phenomenology of each theory at length; our intention is to begin compiling a `dictionary' of theories and their translation into PPF format.

 We will begin with a simple example involving cosmological scalar fields; we progress to more complicated cases which demonstrate how theories involving timelike vector fields, Lorentz violation and brane scenarios can be encapsulated by the PPF parameterization. We then treat Horndeski theory, itself a powerful parameterization that subsumes a large portion of theory space. We conclude with the example of GR supplemented by an exotic fluid. Whilst conceptually simple, this final example has the unusual property that it can be transformed from a theory of two new fields to a single-field theory of either type 1 or type 2.

\subsection{Scalar-Tensor Theory and $f(R)$ Gravity}
\label{subsection:examples_st}
A general scalar-tensor theory has an action of the following form:
\begin{align}
S_{ST}&=\frac{1}{2\kappa}\int \sqrt{-g} \,d^4x \left[\,f(\phi)\, R-K(\phi)\nabla_\mu\phi\nabla^\mu\phi-2 V(\phi)\, \right]\nonumber\\
&+S_M(\psi^a, g_{\mu\nu})
\end{align}
where $\psi^a$ are matter fields. For convenience (in this section only) we have renormalized the scalar field \mbox{$\phi\rightarrow\phi/M_P$} such that it is dimensionless.

In fact one of the functions $f(\phi)$ and $K(\phi)$ is redundant. Through a redefinition of the scalar field one can always write the above action in a  Brans-Dicke-like form, with \mbox{$f(\phi)=\phi$} and \mbox{$K(\phi)=\omega(\phi)/\phi$} \cite{2001PhRvD..63f3504E}. We will work with the action in this form.

As explained in \textsection\ref{subsection:formalism_basics} and Appendix \ref{appendix:fR_transf}, a Legendre transformation maps $f(R)$ gravity into to a scalar-tensor theory, with the following equivalences:
\begin{align}
\phi&\equiv \frac{d\,f(R)}{dR}=f_R & \omega(\phi)&=0 \nonumber\\
 V(\phi)&=\frac{1}{2}\left[R\phi-f(R)\right]
\end{align}
Hence the expressions below can be straightforwardly adapted for use with $f(R)$ models. In fact both theories can be obtained as special cases of the Horndeski Lagrangian (see \textsection\ref{subsection:examples_hd}), but due to their prevalence we will treat them separately here.

The background effective energy density and effective pressure for scalar-tensor theory are (see eqs.(\ref{Friedmann}) and (\ref{Raych})):
\begin{align}
a^2 X=&3\Hu^2(1-\phi)+\frac{1}{2}\omega(\phi)\frac{\dot\phi^2}{\phi}-3{\cal H}\dot\phi+a^2 V(\phi)
 \label{st_background_1}\\
a^2 Y=&-(2\dot\Hu+\Hu^2)(1-\phi)+\frac{1}{2}\omega(\phi)\frac{\dot\phi^2}{\phi}+\ddot\phi+{\cal H}\dot\phi-a^2 V(\phi)
 \label{st_background_2}
\end{align}
The equation of motion for the scalar field is (where matter terms have been eliminated using the gravitational field equations):
\begin{align}
\omega(\phi)\left(\frac{\ddot\phi}{\phi}-\frac{1}{2}\frac{\dot\phi^2}{\phi^2}+2{\cal H}\frac{\dot\phi}{\phi}	\right	)&+\frac{1}{2}\frac{\mathrm{d}\omega}{\mathrm{d}\phi}\frac{\dot\phi^2}{\phi}+a^2\frac{d\,V(\phi)}{d\phi} \nonumber\\*
&-3\left(\dot{\cal H}+{\cal H}^2\right)=0
\label{st_background_eom}
\end{align}

The gauge-invariant combination for perturbations of the scalar field dictated by eqns.(\ref{chi_hat_omega}) and (\ref{omegas}) is:
\begin{equation}
\label{gi_scalar}
\hat\chi=\delta\phi-\frac{\dot\phi}{6\Hu}(\beta+k^2\nu)
\end{equation}
To implement the PPF parameterization, we take the linearly perturbed gravitational field equations of scalar-tensor theory and regroup terms to form the gauge-invariant variables $\hat\Phi$, $\hat\Gamma$ and $\hat\chi$. This puts the equations into the format of eqs.(\ref{FE1}-\ref{FE4}), from which the PPF coefficient functions can be read off:
\begin{widetext}
\begin{align}
 \allowdisplaybreaks
A_0&=-2(1-\phi)+\frac{\dot\phi}{\Hu k^2}\left[\omega\frac{\dot\phi}{\phi}\left(2\Hu+\frac{\dot\Hu}{\Hu}\right)+k^2-6\dot\Hu-a^2\frac{d\,V(\phi)}{d\phi}\right]+3\frac{\ddot\phi}{k^2}&\nonumber\\
B_0&=-\frac{1}{\Hu k}\left[\ddot\phi+\dot\phi\left(\omega\frac{\dot\phi}{\phi}-\Hu-\frac{\dot\Hu}{\Hu}\right)\right] &\nonumber \\
C_0&=2(1-\phi)-\frac{3\phi^{(3)}}{k^2\Hu}-\frac{3\ddot\phi}{k^2\Hu^2}\left(\Hu^2-2\dot\Hu\right)
-\frac{3\dot\phi}{k^2\Hu}\left(4(\dot\Hu+\Hu^2)+\frac{2}{3}k^2-\frac{\omega}{\Hu}\frac{\dot\phi}{\phi}(\dot\Hu+2\Hu^2)-\frac{\ddot\Hu}{\Hu}+\frac{2\dot\Hu^2}{\Hu^2}-a^2\frac{d\,V(\phi)}{d\phi}\right) \nonumber
\end{align}
\vspace{-4mm}
\begin{align}
C_1&=\frac{2}{k\Hu}(1-\phi)\left(k^2+3\Hu^2-3\dot\Hu\right)-\frac{3\dot\phi}{k\Hu^2}(\Hu^2-\dot\Hu)& \nonumber\\
D_0&=1-\phi-\frac{\dot\phi}{\Hu} & D_1&=\frac{k}{\Hu}(1-\phi) \nonumber\\
F_0&=\frac{6}{k}\left[\dot\phi-\Hu (1-\phi)\right]-\frac{\omega \dot\phi^2}{k\Hu\phi} & I_0&=2(1-\phi)-\frac{\dot\phi}{\Hu}\nonumber \\
J_0&=\frac{2}{\Hu k}(1-\phi)(3\dot\Hu+3\Hu^2-k^2)-6\frac{\ddot\phi}{k\Hu}-\frac{3\dot\phi}{k\Hu}\left(2\Hu-\frac{\dot\Hu}{\Hu}+\frac{\omega \dot\phi}{\phi}\right) & J_1&=6(1-\phi)-\frac{3\dot\phi}{\Hu} \nonumber\\
K_0&=-\frac{k}{\Hu}(1-\phi) & K_1&=0 \nonumber\\
\alpha_0&=\frac{1}{k^2}\left[\frac{1}{2}\frac{\mathrm{d}\omega (\phi)}{\mathrm{d}\phi}\,\frac{\dot\phi^2}{\phi}-\frac{1}{2}\omega(\phi)\frac{\dot\phi^2}{\phi^2}-k^2-3 {\cal H}^2+a^2\frac{d\,V(\phi)}{d\phi}\right]  & \alpha_1&=\frac{1}{k}\left[\omega(\phi)\frac{\dot\phi}{\phi}-3{\cal H}\right] \nonumber \\ 
\beta_0&=\frac{1}{k}\left[\omega(\phi)\frac{\dot\phi}{\phi}-{\cal H}\right]  & \beta_1&=1\nonumber\\
\gamma_0&= \frac{3}{k^2}\left[\frac{1}{2} \frac{\mathrm{d}\omega(\phi)}{\mathrm{d}\phi}\,\frac{\dot\phi^2}{\phi}-\frac{1}{2}\omega(\phi)\frac{\dot\phi^2}{\phi^2}+{\cal H}^2+2\dot{\cal H}+\frac{2}{3}k^2-a^2\frac{d\,V(\phi)}{d\phi}\right] \nonumber\\ 
\gamma_1&=\frac{3}{k}\left[{\cal H}+\omega(\phi)\frac{\dot\phi}{\phi}\right] & \gamma_2&=3\nonumber\\
\epsilon_0&=1 & \epsilon_1&=\epsilon_2=0
\end{align}
\end{widetext}
One can verify by direct substitution that these coefficients obey the constraints of eqs.$(\ref{Bianchi_2_constraints_start}-\ref{Bianchi_2_constraints_end})$. Hence scalar-tensor theories and $f(R)$ gravity are type 1 theories according to the classification of \textsection\ref{subsection:formalism_constraints}.

For computations of observables in $f(R)$ gravity it is useful to define a parameter $Q$, which has the rough interpretation as the ratio of the Compton wavelength of the `scalaron' field $f_R$ to the wavelength of a Fourier mode \cite{Pogosian:2007dp}:
\be
Q=3k^2\frac{f_{RR}}{f_R}\approx\left(\frac{\lambda_C}{\lambda}\right)^2
\ee
Making use of the quasistatic approximation (see \textsection\ref{subsection:quasistatic}), the correspondence between our parameterization and $Q$ is given by:
\begin{align}
Q&\approx\frac{3}{2}\left[1-\left(1+D_0-\epsilon_0\frac{B_0}{\beta_0}\right)^{-1}\right]\nonumber\\  
&\approx\frac{3}{2}\left[D_0-\epsilon_0\frac{B_0}{\beta_0}\right]
\end{align}
This result is most easily obtained by mapping both PPF and $f(R)$ gravity onto the common $\{\mu(a,k),\,\gamma(a,k)\}$ parameterization discussed in \textsection\ref{subsection:quasistatic}.

\subsection{Einstein-Aether Theory}
\label{subsection:examples_ae}
Let us now give an example where the new d.o.f. do not arise from a scalar field. Modern Einstein-Aether theory was introduced by Jacobson \& Mattingly \cite{Jacobson:2004ba}  (though in fact the earliest incarnation dates back to Dirac \cite{Dirac, Dirac_el1,Dirac_el2,Dirac_el3}) as a minimalistic model for a theory that violates Lorentz invariance. It achieves this by introducing a dynamical unit vector field, dubbed the `aether', that picks out a preferred spacetime frame. The aether must be time-like at the unperturbed level in order to preserve invariance under spatial rotations, and this requirement is enforced by a Lagrange multiplier in the action \footnote{In this paper we treat only the linear Einstein-Aether theory. In the generalized Einstein-Aether theory ${\cal K}$ is replaced by an arbitrary function ${\cal F}(\cal K)$. This will lead to more complicated PPF coefficients but is not intrinsically problematic.}:
\begin{eqnarray}
\label{EA_action}
S_{EA}&=& \frac{1}{2\kappa^2}\int \mathrm{d}^4x \sqrt{-g}\, \left[R+M^2 {\cal K}+\lambda(A^{\alpha}A_{\alpha}+1)\right]\nonumber\\
&&+S_M[\psi^{a}, g_{\mu\nu}]
\end{eqnarray}
where $M$ has the dimensions of mass and the kinetic term of the aether field is specified by the constants $c_i$:
\begin{eqnarray}
{\cal K}&=&M^{-2} {\cal K} ^{\alpha\beta}_{\gamma\sigma}\nabla_{\alpha}A^{\gamma}\nabla_{\beta}A^{\sigma}\nonumber\\
{\cal K} ^{\alpha\beta}_{\gamma\sigma}&=&c_1 g^{\alpha\beta}g_{\gamma\sigma}+c_2\delta^{\alpha}_{\gamma}\delta^{\beta}_{\sigma}+c_3\delta^{\alpha}_{\sigma}\delta^{\beta}_{\gamma} \label{EA_kappa_tensor}
\end{eqnarray}
For convenience we define \mbox{$\alpha=c_1+3 c_2+c_3$}. The constraint equation from the Lagrange multiplier is \mbox{$g_{\mu\nu}A^\mu A^\nu=-1$}, which leads to \mbox{${A^\mu}=(1/a,0,0,0)$} in an FRW background. 

In our parameterization of the zeroth-order field equations the effective energy density and pressure of the aether are:
\begin{align}
a^2 X&=\frac{3}{2}\alpha\Hu^2 & a^2Y&=-\frac{1}{2}\alpha \left(2\dot\Hu+\Hu^2\right)
\end{align}

The aether experiences perturbations around its zeroth-order direction. We consider only the spin-0 perturbations here, and write the perturbed aether as 
\be
\mathbf{A}=\mathbf{A}^{(0)}+\delta \mathbf{A}=\frac{1}{a}(1+\Xi,\vec{\nabla} V)
\ee
where the perturbed constraint from the Lagrange multiplier has enforced \mbox{$\delta A^0=\Xi/a$}. 

The potential $V$ specifying the perturbation of the spatial aether components represents a new degree of freedom. Following the prescription of \textsection\ref{subsubsection:formalism_dof}, it has the transformation properties $G_4=-1,\,G_3=\Hu$ and $G_1=G_2=0$, so its gauge-invariant partner is: 
\begin{eqnarray}
\hat V&=&V+\frac{1}{6\Hu}(\beta+k^2\nu)-\epsilon\nonumber\\
 &=&V-\frac{\hat\Phi}{\Hu}+\frac{1}{2}\dot\nu
\end{eqnarray}
$\hat V$ has the dimension of a length, so a factor of $k$ is included when defining the dimensionless perturbation appearing in the parameterization of eqs.(\ref{FE1}--\ref{FE4}), i.e. $\hat\chi\equiv k\hat V$.

With the preparations completed, we can now put the linearized field equations of Einstein-Aether theory into the standard format of our parameterization. The PPF coefficients can then be read off:
\begin{align}
 \allowdisplaybreaks
A_0 &=c_1\left(1-\frac{\dot\Hu}{\Hu^2}\right)-\alpha & B_0 &=\frac{k}{\Hu}(c_1+c_2+c_3)\nonumber\\
C_0 &=\alpha\left(2-\frac{\dot\Hu}{\Hu^2}\right) & C_1 &=\frac{\alpha}{k\Hu}\left(k^2+3\Hu^2-3\dot\Hu\right)\nonumber\\
D_0 &=-(c_1+c_3)\left(2-\frac{\dot\Hu}{\Hu^2}\right) & D_1 &=-\frac{k}{\Hu}(c_1+c_3)\nonumber\\
F_0 &=\frac{k}{\Hu}(c_1-3\alpha\Hu_k^2) & I_0 &=\alpha\nonumber\\
J_0 &=\frac{3\alpha}{k\Hu}\left(\Hu^2+\dot\Hu\right) & J_1 &=3\alpha\nonumber\\
K_0 &=0 & K_1 &=0\nonumber\\
\alpha_0&=\Hu_k(\alpha-c_1) & \alpha_1&=c_1 \nonumber \\
\beta_0&=(c_1+c_2+c_3) & \beta_1&=0 \nonumber \\
\gamma_0&=2\Hu_k\alpha & \gamma_1&=\alpha \nonumber\\
\gamma_2&=0 & \epsilon_0&=-2\Hu_k(c_1+c_3) \nonumber\\
\epsilon_1&=-(c_1+c_3) & \epsilon_2&=0
\end{align}
These coefficients satisfy the constraints of eqs.(\ref{Bianchi_2_constraints_start}-\ref{Bianchi_2_constraints_end}), hence Einstein-Aether is a type 1 theory.

\subsection{DGP}
\label{subsection:examples_dgp}
Dvali-Gabadadze-Porrati gravity (DGP) \cite{2000PhLB..485..208D} is undoubtedly the most commonly-discussed member of the braneworld class of modified gravity theories. It is also one of the most extensively tested: recent results \cite{Raccanelli2012} have placed tight constraints on the normal (non-accelerating) branch of the theory in addition to the existing constraints on the self-accelerating branch \cite{Fang:2008kc}. Although increasingly disfavoured, we will present its PPF correspondence here as a representative example of braneworld scenarios \cite{Maartens:2010to}.

The theory considers our four-dimensional spacetime to be embedded in a five-dimensional bulk, with standard matter fields confined to the 4D brane. The ratio of the effective gravitational constants in the bulk and brane defines a crossover scale, \mbox{$r_c=\kappa_5/(2 \kappa_4)$} (constrained to be of order the horizon scale by supernovae data), above which gravitational forces are sensitive to the additional dimension. Below the crossover scale the theory can be treated as effectively four-dimensional. However, this is not to say that DGP reduces to GR below the crossover scale; the theory allows a new scalar d.o.f. to propagate \footnote{The new scalar d.o.f. can arise from either a brane-bending mode (if the brane tension is negative), the helicity-0 mode of the graviton (if the brane tension is positive), or a superposition of the two (for zero brane tension) \cite{Koyama2005, Gorbunov2006}.},
 and enters a scalar-tensor-like regime before strong coupling of the scalar switches off departures from GR at the Vainshtein scale \cite{Vainshtein1972}. 

Neglecting brane tension, the full action of the theory is:
\begin{eqnarray}
\label{DGP_action} 
S_{DGP} &=& \frac{1}{2\kappa_5}\int d^5\tilde{x} \sqrt{-^{(5)}\tilde{g}}\, \left[^{(5)}\tilde{R}-2\Lambda_5\right]\\
&&+\,\int d^4x\, \sqrt{-^{(4)}g}\left[\frac{1}{2\kappa_4}{^{(4)}R}+{\cal L}_M (\psi^a, g_{\mu\nu})\right] \nonumber
\end{eqnarray}
${\cal L}_M$ is the Lagrangian of the brane-confined matter fields and $^{(5)}\tilde{R}$ is the Ricci scalar of the 5D spacetime metric $^{(5)}\tilde{g}_{\mu\nu}$. The effective energy density,  pressure and equation of state that modify the Friedmann and Raychaudhuri equations are (see eqs.(\ref{Friedmann}) and (\ref{Raych})):
\begin{align}
X&=\frac{3\varepsilon}{a\,r_c}\Hu\\
Y&=-\frac{\varepsilon}{a\,r_c \Hu}(\dot\Hu+2\Hu^2)\\
 \omega_E&=\frac{Y}{X}=-\frac{1}{3}\left(\frac{\dot\Hu}{\Hu^2}+2\right)\label{omega_eq}
\end{align}
$\varepsilon=+1$ corresponds to the self-accelerating solution branch, which is asymptotically de Sitter at late times but suffers from ghost pathologies \cite{Charmousis2006,Gorbunov2006}. $\varepsilon=-1$ corresponds to the normal solution branch, on which a cosmological constant or dark energy is still required to achieve accelerated expansion. 

The new fields in DGP arise from the projection of the electric part of the bulk Weyl tensor onto the brane. It is common to treat the components of the projected Weyl tensor $E_{\mu\nu}$ as an effective `Weyl fluid' with a radiation-like equation of state, $\omega_W=1/3$. The subdominance of the Weyl terms at late times means that the Weyl fluid is usually defined at the perturbative level, and its zeroth-order components are neglected. The effective Weyl fluid perturbations are:
\begin{align}
a^2 E^0_0&=\kappa_4 a^2 \rho_W \delta_W  &\\
a^2 E^0_i&=\nabla_i \left[\kappa_4 a^2 \rho_W (1+\omega_W)\theta_W \right]\nonumber\\
a^2 E^i_j&=-\left(\frac{1}{3}\delta^i_j\,\kappa_4a^2 \rho_W\delta_W+D^i_j \left[\kappa_4 a^2 \rho_W (1+\omega_W)\Sigma_W\right]\right)\nonumber
 \end{align}
There are three new fields corresponding to the energy density perturbation, velocity potential and anisotropic stress of the Weyl fluid, so the two components of the four-dimensional $U$-conservation law do not provide us with sufficient equations of motion to solve the system. By considering perturbations of the full five-dimensional spacetime one can relate the three Weyl perturbations to a single d.o.f., the master variable \cite{Mukohyama:2000cr}. However, solution of the e.o.m. for the master variable requires gradients perpendicular to the brane that are not encapsulated by the four-dimensional formalism of this paper. Hence we must stick with three new fields, and thus DGP is not subject to the classification of \textsection\ref{subsection:formalism_constraints}. 

The gauge-invariant Weyl fluid variables are:
\begin{align}
 \hat\delta_W&=\delta_W+\frac{1+\omega_W}{2}(\beta+k^2\nu) \\
 \hat\theta_W&=\theta_W-\frac{1}{6\Hu}(\beta+k^2\nu) \nonumber\\
 \hat{\Sigma}_W&=\Sigma_W\nonumber
\end{align}
 The metric PPF coefficients for DGP are:
 \begin{align}
  \allowdisplaybreaks
A_0&=-\frac{3}{r_c^2 X} \left(1+\frac{3\,\kappa a^2\rho_W (1+\omega_W)}{2k^2}\right) & \nonumber\\
B_0&=\frac{3}{r_c^2 X}\frac{\kappa a^2\rho_W (1+\omega_W)}{2k\Hu}& \nonumber\\
C_0&=\frac{3}{r_c^2 X}\left[4+3\,\omega_E+\frac{3\,\kappa a^2\rho_W (1+\omega_W)}{2 k^2}\left(2+3\omega_E\right)\right] & \nonumber\\
C_1&=\frac{3}{r_c^2 X}\frac{1}{\Hu k}(k^2+3\Hu^2-3\dot\Hu)\nonumber\\
D_0&=-\frac{3}{r_c^2 (X+3Y)} &\nonumber\\
D_1&=-\frac{k}{\Hu}\frac{3}{r_c^2 (X+3Y)}\nonumber\\
F_0&=-\frac{9\Hu_k}{r_c^2 X}\nonumber\\
I_0&=\frac{3}{r_c^2 X}\nonumber\\
J_0&=\frac{3}{r_c^2 X}\frac{1}{k \Hu}\left( -k^2+3\Hu^2(4+3\omega_E)+3\dot\Hu\right) &\nonumber\\
J_1&=\frac{9}{r_c^2 X}\nonumber\\
K_0&=\frac{k}{\Hu}\frac{3}{r_c^2 (X+3Y)}\nonumber  \\
K_1&=0
\end{align}
To incorporate three new fields, in principle the template of eqs.(\ref{FE1}-\ref{FE4}) needs to be extended to include two more sets of terms similar to the $\hat\chi$'s, each with their own coefficient functions. However, it turns out that for DGP most of these possible extra terms are not needed, and the only non-zero $\hat\chi$-coefficients are:
\begin{align}
\alpha_0^{\hat\delta}&=-\frac{3}{2Xr_c^2}\frac{\kappa a^2\rho_W }{k^2} \nonumber\\
\beta_0^{k\hat\theta}&=-\frac{3}{Xr_c^2}\frac{\kappa a^2\rho_W (1+\omega_W)}{2k^2}\nonumber \\
\gamma_0^{\hat\delta}&=\frac{3}{Xr_c^2}\frac{3\,\kappa a^2\rho_W}{2k^2}\left(2+3\omega_E\right)\nonumber\\
\epsilon_0^{k^2\hat\Sigma}&=\frac{3}{r_c^2 (X+3Y)} \frac{\kappa a^2\rho_W (1+\omega_W)}{k^2}
\end{align}
where the superscripts indicate the degree of freedom to which each coefficient belongs, and should not be confused with spacetime indices. Note that, as per the previous subsection, factors of $k$ have been used where necessary to define dimensionless perturbations of the Weyl fluid.

The authors of \cite{Wagoner:2008vg} reported that a distinctive signature of DGP is the unusual scale-dependence of new terms in the modified Poisson equation. Specifically, they found that DGP led to terms linear in wavenumber $k$ as the result of a brane-bending mode. 
PPF is formulated in four dimensions, so off-brane terms such as this that explicitly probe the additional dimension are not present in our coefficient functions. In \textsection\ref{subsection:scale_dep} we will see that the four-dimensional theory shares the $k$-dependence of a scalar-tensor theory.

\subsection{Eddington-Born-Infeld Gravity}
\label{subsection:examples_ebi}

Let us give another demonstration of how the PPF parameterization can handle theories beyond scalar-field-type models, this time by considering a bimetric theory. Bimetric theories have been the subject of recent intense interest in the context of Massive Gravity (see \cite{Hinterbichler2012} for a review); we will instead focus on the related theory of Eddington-Born-Infeld (EBI) gravity, where the coupling between the two metrics is substantially simpler. EBI gravity possesses a second graviton, but unlike Massive Gravity the associated scalar mode propagates and could potentially be a ghost.

EBI gravity arose from the realisation of \cite{Banados:2008bc} that a theory coupling GR to a Born-Infeld-type Lagrangian could be reformulated as an action for two coupled metrics (a closely-related theory based on a bimetric reformulation of Eddington's original affine action \cite{Eddington} was explored in \cite{Banados2010, Celia2012, AvelinoFerreira2012, Casanellas:2011vh, Sham_EBI_2012}). The bimetric re-writing of the EBI action is: 
\be
S_{EBI}&=&\frac{1}{16\pi G}\int d^4x\Big[\sqrt{-g}(R-2\Lambda)+\sqrt{-\tilde{q}}(\tilde{K}-2\tilde\lambda)\nonumber\\
&&-\frac{\sqrt{-\tilde{q}}}{l^2}(\tilde{q}^{-1})^{\mu\nu}g_{\mu\nu}\Big]+S_M[\psi^a, g_{\mu\nu}]
\label{ebi_bimetric_action}
\ee
Matter couples to the usual spacetime metric $g_{\mu\nu}$ and $\tilde K$ is the curvature of the auxiliary metric $\tilde{q}_{\mu\nu}$. $\tilde\lambda$ can be considered as a cosmological constant for the auxiliary metric and $l$ is a length.

For the purpose of the PPF framework we require the cosmological perturbation theory laid out in \cite{Banados:2009ec}. First note that, since our conformal time coordinate is defined by the standard spacetime metric, the `scale factors' of the temporal and spatial parts of the auxiliary metric are not constrained to be equal to oneanother nor the cosmological scale factor $a(\eta)$. Instead we write the unperturbed auxiliary metric as:
\begin{align}
\tilde{q}_{00}&=-Z(\eta)^2 & \tilde{q}_{ij}&=R(\eta)^2\gamma_{ij}  
\end{align}
The background-level effective energy density and effective pressure are:
\begin{align}
a^2X&=\kappa a^2\rho_{EBI}=\frac{R^3}{l^2 Z a} \\
a^2Y&=\kappa a^2 P_{EBI}=-\frac{aRZ}{l^2} 
\end{align}
The scalar perturbations of the auxiliary metric are written in an analogous manner to those of the spacetime metric in eq.(\ref{line_element}):
\be
\label{EBI_line_element}
ds^2 &=& -Z(\eta)^2 (1-2 \tilde{\Xi})d\eta^2-2R(\eta)^2\,(\vec{\nabla}_i\tilde{\epsilon})d\eta\,dx\nonumber\\
&+&R(\eta)^2\left[\left(1+\frac{1}{3}\tilde\beta\right)\gamma_{ij}+D_{ij}\tilde\nu\right] dx^i\,dx^j 
\ee
However, we need to be aware that the auxiliary metric perturbations transform in a slightly different way to those of the spacetime metric, due to the possible inequality of $Z(\eta)$ and $R(\eta)$. We must use the same gauge transformation vector as for the spacetime metric, \mbox{$\xi^\mu=1/a (\xi^0,\nabla^i\psi)$}. The auxiliary perturbations then transform as (using primes to denote the transformed variables):
\begin{align}
\tilde\Xi^{\prime}&=\tilde\Xi-\frac{1}{a}\left[\left(\frac{\dot Z}{Z}-\Hu\right){\xi^0}+\dot{\xi^0}\right]\nonumber\\
\tilde\epsilon^{\prime}&=\tilde\epsilon+\frac{1}{a}\left[\frac{a^2 Z^2}{R^2}{\xi^0}+\Hu\psi-\dot\psi\right]\nonumber\\
\tilde\beta^{\prime}&=\tilde\beta+\frac{1}{a}\left[\frac{6\dot R}{R}{\xi^0}-2k^2\psi\right]\nonumber\\
\tilde\nu^{\prime}&=\tilde\nu+\frac{2\psi}{a}
\end{align}
These transformations reduce to those of the ordinary spacetime metric perturbations for the case \mbox{$Z(\eta) = R(\eta) = a(\eta)$}. The four scalar perturbations of the auxiliary metric are the new fields of EBI gravity. Using the now-familiar algorithm of \textsection\ref{subsection:formalism_vars}, the corresponding gauge-invariant combinations are:
\begin{align}
\hat{\tilde\Xi}&=\tilde\Xi-\Xi -\frac{1}{6\Hu}\left(\Hu-\frac{\dot Z}{Z}\right)(\beta+k^2\nu)\nonumber\\
\hat{\tilde\epsilon}&=\tilde\epsilon-\epsilon -\frac{1}{6\Hu}\left(\frac{Z^2}{R^2}-1\right)(\beta+k^2\nu)\nonumber\\
\hat{\tilde\beta}&=\tilde\beta+k^2\nu-\frac{\dot R}{\Hu R}(\beta+k^2\nu)\nonumber \\
\hat{\tilde\nu}&=\tilde\nu-\nu 
\end{align}
We put the linearized field equations of the spacetime metric into the standard format of eqs.(\ref{FE1}-\ref{FE4}), with each of the gauge-invariant auxiliary perturbations $\hat{\tilde\Xi},\,\hat{\tilde\epsilon},\,\hat{\tilde\beta}$ and $\hat{\tilde\nu}$ acting as a new $\hat\chi$-type field. The PPF coefficients are (where the superscripts on the Greek coefficients indicate to which field they belong):
\begin{align}
A_0&=-\frac{R^3}{\Hu k^2 l^2 Z a}\left(\frac{3\dot R}{R}-\frac{\dot Z}{Z}-2\Hu\right)\nonumber\\
B_0&=\frac{R^3}{k^2 l^2 Z a}\frac{k}{\Hu}\left(\frac{Z^2}{R^2}-1\right)\nonumber\\
C_0&=\frac{3ZRa}{\Hu k^2 l^2}\left(\frac{\dot R}{R}+\frac{\dot Z}{Z}-2\Hu\right)
\end{align}
\begin{align}
\alpha_0^{\tilde\Xi}&=\frac{R^3}{k^2 l^2 Z a} & \alpha_0^{\tilde\beta}&=\frac{R^3}{2k^2 l^2 Z a} \nonumber\\
\beta_0^{k{\tilde\epsilon}}&=-\frac{R^3}{k^2 l^2 Z a} \nonumber\\
\gamma_0^{\tilde\Xi}&=\frac{3ZR a}{k^2 l^2} & \gamma_0^{\tilde\beta}&=-\frac{ZR a}{2k^2 l^2} \nonumber\\
\epsilon_0^{k^2{\tilde\nu}}&=-\frac{ZR a}{k^2 l^2} 
\end{align}
Factors of $k$ are included where necessary to define dimensionless perturbations to the auxiliary metric. 
The four additional fields present in EBI gravity means it is not subject to our type 1/2/3 classification. In addition to both components of the $U$-conservation law, two further relations are required to evolve the system; these are provided by eqs.(33) and (34) of \cite{Banados:2009ec}.

\subsection{Ho\u{r}ava-Lifschitz Theory}
\label{subsection:examples_hl}

Ho\u{r}ava-Lifschitz gravity \cite{Horava2009, Sotiriou_HL_review} was proposed as a theory of quantum gravity which achieves power-counting renormalizability at ultra-violet energy scales, whilst potentially recovering GR at low energies. GR is rendered non-renormalizable by the scaling properties of the graviton propagator, but this situation can be remedied by introducing higher-order derivative terms in the gravitational action \cite{Kimpton:2010uy}. However, the price paid for this improved UV behaviour is the appearance of a ghostly (negative energy) degree of freedom. The origin of the ghost can be traced to the presence of higher-order \textit{time} derivatives in the action.

Ho\u{r}ava's theory sidesteps this issue by breaking Lorentz invariance and treating time and space unequally. The Ho\u{r}ava-Lifschitz action contains nonlinear spatial curvature terms that introduce greater-than-second-order spatial derivatives without introducing higher time derivatives. Consequently the full diffeomorphism group of GR is broken, and only the smaller group of foliation-preserving diffeomorphisms is maintained. As is expected, the breaking of general covariance introduces a new scalar graviton mode; there has been considerable effort made to determine whether this mode propagates or not \cite{2009JHEP...08..070C,2009JHEP...10..029B,2010PhRvD..81h3508G}. 

After using the ADM formalism to define a preferred time direction, two flavours of Ho\u{r}ava-Lifschitz gravity exist depending on whether the lapse function is allowed to be a function of both space and time (the \textit{non-projectable} theory) or time only (the \textit{projectable} theory). In the non-projectable theory the action picks up extra terms that depend on the shift vector via:
\begin{equation}
b_i=\partial_i \mathrm{ln}\, N(t,\vec{x})
\end{equation}
where $N$ is the lapse function and we avoid the standard `$a_i$' notation to prevent confusion with the cosmological scale factor.

 We will focus on the non-projectable extension of Ho\u{r}ava's original action put forward by Blas, Pujol\`as and Sibiryakov \cite{2010PhRvL.104r1302B}, which has dominated much of the recent discussion. The Blas \textit{et al.} theory is motivated by the requirement that the new scalar mode possesses a healthy quadratic kinetic term, and also lifts the `detailed balance' condition of Ho\u{r}ava's original theory. The Blas \textit{et al.} action is:
\begin{align}
S_{HL}=&\frac{M_P^2}{2}\int\, dt\, d^3x \,\sqrt{-g}\,N\left[{\cal L}_K-V(g_{ij}, b_i)\right]\nonumber\\
&+\int\, dt\, d^3x \,\sqrt{-g}\,N {\cal L}_M(N, N_i, g_{ij}) \\
\mathrm{where}\quad\quad{\cal L}_K&=K_{ij} K^{ij}-\lambda K^2\nonumber\\
K_{ij}&=\frac{1}{2N}\left(\partial_t g_{ij}-\nabla_iN_j-\nabla_jN_i\right)\nonumber\\
V(g_{ij}, b_i)=&-R-c\, b_i b^i+M_P^{-2}(d_2 R^2+d_3R_{ij}R^{ij}\nonumber\\
&+c_2 b_i\Delta b^i+c_3 R\nabla_i b^i+...)\nonumber\\
&+M_P^{-4}(d_4 R\Delta R+d_5\nabla_i R_{jk}\nabla^iR^{jk}\nonumber\\
&+c_4b_i\Delta^2 b^i+c_5\Delta R\nabla_i b^i+...).
\label{hl_action}
\end{align}
Ellipses indicate terms not relevant for linear spin-0 perturbations, and $c_i$ and $d_i$ are constants. 
The recovery of GR in the limit $\lambda=1$ is not clear-cut, as the new scalar mode becomes strongly coupled at low energies. Resolution of the issue depends on whether the scalar mode propagates, see the references above.

In a flat FRW spacetime the zeroth-order field equations of Ho\u{r}ava-Lifschitz gravity are identical to those of GR but with a rescaled gravitational constant, \mbox{$G_{\mathrm{eff}}= 2 G_0 /(3\lambda-1)$}. In terms of our background-level parameterization this becomes:
\begin{align}
a^2X &=\kappa a^2\rho_M\frac{3(1-\lambda)}{3\lambda-1}=3(1-\lambda) \frac{3\Hu^2}{2}\\
 a^2Y &=\kappa a^2 P_M\frac{3(1-\lambda)}{3\lambda-1}=-\frac{3}{2}(1-\lambda)(2\dot\Hu+\Hu^2)
\end{align}
Moving on to parameterization of the linearized field equations, we need to be aware of the reduced diffeomorphism group of Ho\u{r}ava-Lifschitz gravity. The theory is invariant under the gauge transformations \mbox{$\eta\rightarrow \tilde{\eta}(\eta),\, x^i\rightarrow\tilde{x}^i(\eta,\vec{x})$}, i.e. we have lost the ability to perform space-dependent reparameterizations of the time coordinate. However, by making use of the St\"{u}ckelberg trick we can restore full general covariance and remove the graviton scalar mode from the metric sector of the theory; instead we can recast it as a extra field, thereby obtaining something similar to a scalar-tensor theory.

The St\"{u}ckelberg trick is implemented by promoting the temporal component of the gauge transformation vector \mbox{$\xi^\mu=1/a({\xi^0},\nabla^i\psi)$} to a new field possessing the necessary transformation properties to ensure gauge invariance of the whole equation. Once these transformation properties are determined we can use our standard procedure of \textsection\ref{subsubsection:formalism_dof} to construct the gauge-invariant variable for the new St\"{u}ckelberg field. Note that we are really using a quasi-St\"{u}ckelberg trick, as we are leaving the spatial diffeomorphisms unaltered.

Let the dimensionless St\"{u}ckelberg perturbation be \mbox{$\chi=k\,{\xi^0}$} (${\xi^0}$ has the dimension of length). The accompanying gauge-invariant variable for the St\"{u}ckelberg field is:
\be
\hat\chi=\chi+\frac{(\beta+k^2\nu)}{6\Hu_k}
\ee
For convenience we also define the following higher-derivative spatial operators:
\begin{align} 
f_1&=-\left(\frac{2c_3}{M_P^2}\frac{k^2}{a^2}+\frac{2c_5}{M_P^4}\frac{k^4}{a^4}\right) \\
f_2&=-\left(c+\frac{c_2}{M_P^2}\frac{k^2}{a^2}+\frac{c_4}{M_P^4}\frac{k^4}{a^4}\right)\\
f_3&=-2\left(\frac{(8d_2+3d_3)}{M_P^2}\frac{k^2}{a^2}+\frac{(8d_4-3d_5)}{M_P^4}\frac{k^4}{a^4}\right) & 
\end{align}
The PPF coefficient functions are found to be:
\begin{align}
 \allowdisplaybreaks
A_0&=3(\lambda-1)+f_2\left(1-\frac{\dot\Hu}{\Hu^2}\right) \nonumber\\
B_0&=\frac{k}{\Hu}(1-\lambda) \nonumber\\
C_0&=3(1-\lambda)\left(2-\frac{\dot\Hu}{\Hu^2}\right) -f_1\left(1-\frac{\dot\Hu}{\Hu^2}\right)\nonumber\\
C_1&=\frac{3}{\Hu k}(1-\lambda)\left[3(\Hu^2-\dot\Hu)+k^2\right] \nonumber
\end{align}
\begin{align}
D_0&=-\frac{f_1}{2}\left(1-\frac{\dot\Hu}{\Hu^2}\right) & D_1&=0 \nonumber\\
F_0&=9\Hu_k (1-\lambda)+\frac{k}{\Hu}f_2 & I_0&=3(1-\lambda) \nonumber
\end{align}
\vspace{-4mm}
\begin{align}
J_0&=\frac{1}{\Hu k}\left[9(1-\lambda)(\Hu^2+\dot\Hu)-k^2 f_1\right]\nonumber\\
 J_1&=9(1-\lambda) \nonumber
\end{align}
\vspace{-6mm}
\begin{align}
K_0&=-\frac{k}{\Hu}\frac{f_1}{2} & K_1&=0 \nonumber\\
\alpha_0&=\Hu_k \left[3(\lambda-1)+f_1+f_2\right] & \alpha_1&=f_2 \nonumber\\
\beta_0&=1-\lambda & \beta_1&=0 \nonumber\\
\gamma_0&=\Hu_k \left[6(1-\lambda)-f_1+f_3\right] & \gamma_1&=3(1-\lambda)-f_1\nonumber \\
\epsilon_0&=\frac{1}{2}\Hu_k (f_3-f_1) & \epsilon_1&= -\frac{f_1}{2}
\end{align}
The Ho\u{r}ava-Lifschitz coefficients obey the constraint relations of eqs.(\ref{Bianchi_2_constraints_start}-\ref{Bianchi_2_constraints_end}), hence it is classified as a type 1 theory. 

We see that Ho\u{r}ava-Lifschitz theory possesses a distinctive signature: nearly all of its PPF coefficients contain spatial derivative operators that are greater than second order. This feature is a result of the particular type of Lorentz violation of Ho\u{r}ava-Lifschitz theory, which allows time and space coordinates to behave differently under scaling transformations in the UV. 
 
It is these kinds of distinctive signatures that could be a useful tool in guiding us towards (or eliminating) particular regions of theory space when confronted with data. In the case of Ho\u{r}ava-Lifschitz gravity, the challenge will be the detection of strongly scale-dependent components of the PPF functions which are likely to be subdominant to the `normal' scale-free and $k^2$ terms (see \textsection\ref{subsection:scale_dep}).

\subsection{Horndeski Theory}
\label{subsection:examples_hd}

The last few years have witnessed a resurgence of interest in Horndeski theory \cite{Horndeski:1974ko}, which had lain largely forgotten since 1974 until it was independently re-derived by Deffayet and collaborators \cite{Deffayetetal_2011}. Horndeski theory is the most general Lorentz-invariant extension of GR in four dimensions that can be constructed using a single additional scalar field, with the restriction that the equations of motion must remain second order in time derivatives. All theories that fit this description can be obtained via special choices of four arbitrary functions of the scalar field that appear in the Horndeski Lagrangian. A non-exhaustive list of theories that fall under the Horndeski umbrella is: Brans-Dicke and scalar-tensor gravity, $f(R)$ gravity (in its scalar-tensor formulation), single-field quintessence and K-essence theories, single-field inflation models, the covariant Galileon, the Fab Four, Dirac-Born-Infeld theory, Kinetic Gravity Braiding, actions involving derivative couplings between a scalar field and the Einstein tensor and $f({\cal G})$ theories, where $\cal G$ is the Gauss-Bonnet term.
\begin{center}
\begin{table*}
\renewcommand{\arraystretch}{1.75}
\begin{tabular}{| c | c | c | c | c |} \hline
\bf{Theory} & $\bf{K(\phi, {\cal X})}$ & $\bf{G_3(\phi, {\cal X})}$ & $\bf{G_4(\phi, {\cal X})}$ & $\bf{G_5(\phi, {\cal X})}$ \\ \hline
Scalar-tensor theory & $\frac{M_P\omega (\phi) {\cal X}}{\phi}-V(\phi)$ & 0 & $\frac{M_P}{2}\phi$ & 0 \\ \hline
$f(R)$ gravity & $-\frac{M_P^2}{2}(R f_R-f(R))$ & 0 & $\frac{M_P^2}{2} f_R$ & 0 \\ \hline
the covariant Galileon & $-c_{2}{\cal X}$ & $\frac{c_{3}}{m^{3}}\, {\cal X}$ & $\frac{M_P^2}{2}-\frac{c_{4}}{m^{6}}\, {\cal X}^{2}$ & $\frac{3c_{5}}{m^{9}}\, {\cal X}^{2}$\nonumber\\ \hline
{Horndeski's original notation} & \multirow{2}{*}{$\kappa_9+4{\cal X}\left(\breve{\kappa}_{8,\phi}-2\breve{\kappa}_{3,\phi\phi}\right)$}
& $-2\big(6\breve{\kappa}_{1,\phi\phi}+8\breve{\kappa}_{1,\phi}$
 &\multirow{2}{*}{$-4\left(\breve{\kappa}_{1,\phi}+\breve{\kappa}_{3}-{\cal X}\breve{\kappa}_{3,{\cal X}}\right)$} & \multirow{2}{*}{$-4\kappa_1$}  \\ 
 (used for the Fab Four) & &$-\breve{\kappa}_8+{\cal X}\kappa_8-8{\cal X}\kappa_{3,\phi}\big)$ & & \\ \hline
Kinetic Gravity Braiding & $K(\phi, {\cal X})$ & $G_3(\phi, {\cal X})$ & $\frac{1}{2}M_P^2$ & 0 \\ \hline
Quintessence \& phantom fields & ${\varepsilon}{\cal X}-V(\phi)$ & 0 & $\frac{1}{2}M_P^2$ & 0 \\ \hline
K-essence \& K-inflation & $K(\phi, {\cal X})$ & 0 & $\frac{1}{2}M_P^2$ & 0 \\ \hline
\end{tabular}
\label{table:HD}
\caption{This table lists the choices for the four free functions of the Horndeski Lagrangian that reproduce some previously-studied theories of modified gravity. In the table above, if a function is left in general terms then the choice is arbitrary. For the covariant Galileon, $c_i$ are dimensionless constants and $m$ is a mass scale. For the fourth line, \mbox{$\breve{\kappa}_i^=\int \kappa_i\, d{\cal X}$} and expressions for the $\kappa_i$ that give rise to the Fab Four are presented in \cite{Charmousis2012}. In the penultimate line, \mbox{$\varepsilon=+1$} for quintessence and \mbox{$\varepsilon=-1$} for phantom scalar fields.} 
\end{table*}
\end{center}

Inevitably, the price to be paid for this powerful generality is that Horndeski theory is cumbersome to calculate with. Its action is \cite{Gao_Steer_2011,DeFelice:2011uq}:
 \begin{equation}
S_{HD}=\int d^{4}x\sqrt{-g}\left(\sum_{i=2}^{5}{\cal L}_{i}+{\cal L}_{M}(\psi^a,g_{\mu\nu})\right)\,,\label{hd_action}
\end{equation}
 where 
\begin{widetext}
\begin{eqnarray}
{\cal L}_{2} & = & K(\phi,{\cal X})\label{eachlag2}\\
{\cal L}_{3} & = & -G_{3}(\phi,{\cal X})\Box\phi\\
{\cal L}_{4} & = & G_{4}(\phi,{\cal X})\, R+G_{4,{\cal X}}\,[(\Box\phi)^{2}-(\nabla_{\mu}\nabla_{\nu}\phi)\,(\nabla^{\mu}\nabla^{\nu}\phi)]\\
{\cal L}_{5} & = & G_{5}(\phi,{\cal X})\, G_{\mu\nu}\,(\nabla^{\mu}\nabla^{\nu}\phi)-\frac{1}{6}\, G_{5,{\cal X}}\,[(\Box\phi)^{3}-3(\Box\phi)\,(\nabla_{\mu}\nabla_{\nu}\phi)\,(\nabla^{\mu}\nabla^{\nu}\phi)+2(\nabla^{\mu}\nabla_{\alpha}\phi)\,(\nabla^{\alpha}\nabla_{\beta}\phi)\,(\nabla^{\beta}\nabla_{\mu}\phi)]\label{eachlag5}
\end{eqnarray}
$K$ and $G_i$ (i=3,4,5) are four functions of a scalar field and its kinetic energy,  ${\cal {\cal X}}=-\partial^{\mu}\phi\partial_{\mu}\phi/2$. GR is recovered by setting $G_4=M_P^2/2$ and $G_3=G_5=K=0$. The cosmological perturbations of the Horndeski action were first derived in \cite{Gao_Steer_2011}, but we will follow the notation of \cite{DeFelice:2011uq} by defining the following useful quantities :
\begin{align}
 \allowdisplaybreaks
\FT & \equiv  2\left[G_{4}-{\cal X}\left(\frac{\ddot{\phi}}{a^2}\, G_{5,{\cal X}}-\frac{\Hu \dot\phi}{a^2}\, G_{5,{\cal X}}+G_{5,\phi}\right)\right]\\[5pt]
\GT & \equiv 2\left[G_{4}-2{\cal X}G_{4,{\cal X}}-{\cal X}\left(\frac{\Hu\dot{\phi}}{a^2}\, G_{5,{\cal X}}-G_{5,\phi}\right)\right]\\[5pt]
a\,\Theta 
 & =  -\dot{\phi}{\cal X}G_{3,{\cal X}}+2\Hu G_{4}-8\Hu {\cal X}G_{4,{\cal X}}-8\Hu {\cal X}^{2}G_{4,{\cal X}{\cal X}}+\dot{\phi}G_{4,\phi}+2{\cal X}\dot{\phi}\, G_{4,\phi {\cal X}}\nonumber \\
 &  -\frac{\Hu^{2}}{a^2}\dot{\phi}\left(5{\cal X}G_{5,{\cal X}}+2{\cal X}^{2}G_{5,{\cal X}{\cal X}}\right)+2\Hu {\cal X}\left(3G_{5,\phi}+2{\cal X}G_{5,\phi {\cal X}}\right)\\[5pt]
a^2\,\Upsilon & =  a^2\{{\cal X}K_{,{\cal X}}+2{\cal X}^{2}K_{,{\cal X}{\cal X}}\}+12\Hu\dot{\phi}{\cal X}G_{3,{\cal X}}+6\Hu\dot{\phi}{\cal X}^{2}G_{3,{\cal X}{\cal X}}-2a^2\{{\cal X}G_{3,\phi}+{\cal X}^{2}G_{3,\phi {\cal X}}\}\nonumber \\
 &   -6\Hu^2 G_{4}+6\left[\Hu^{2}\left(7{\cal X}G_{4,{\cal X}}+16{\cal X}^{2}G_{4,{\cal X}{\cal X}}+4{\cal X}^{3}G_{4,{\cal X}{\cal X}{\cal X}}\right)-\Hu\dot{\phi}\left(G_{4,\phi}+5{\cal X}G_{4,\phi {\cal X}}+2{\cal X}^{2}G_{4,\phi {\cal X}{\cal X}}\right)\right]\nonumber \\
 &   +\frac{1}{a^2}\left\{30\Hu^{3}\dot{\phi}{\cal X}G_{5,{\cal X}}+26\Hu^{3}\dot{\phi}{\cal X}^{2}G_{5,{\cal X}{\cal X}}+4\Hu^{3}\dot{\phi}{\cal X}^{3}G_{5,{\cal X}{\cal X}{\cal X}}\right\}-6\Hu^{2}{\cal X}\bigl(6G_{5,\phi}+9{\cal X}G_{5,\phi {\cal X}}+2{\cal X}^{2}G_{5,\phi {\cal X}{\cal X}}\bigr)
\end{align}
 where the relation ${\cal X}\partial_{{\cal X}}\dot{\phi}=\dot{\phi}/2$ has been used. Note that we have used conformal time, contrary to the authors of \cite{DeFelice:2011uq}. The gauge-invariant perturbation of the scalar field is:
 \begin{equation}
\label{gi_scalar2}
\hat\chi=\frac{\delta\phi}{M_{P}}-\frac{\dot\phi}{M_{P}}\frac{1}{6\Hu}(\beta+k^2\nu)
\end{equation} 
 Using a tilde to denote division by the square of the reduced Planck mass (ie. \mbox{$\GTT=\GT/M_{P}^2=\kappa\,\GT$}), the PPF coefficients for Horndeski theory are:
 \begin{align}
 \allowdisplaybreaks
A_0 &=-2\left(1-\frac{a\tilde\Theta}{\Hu}\right)-\frac{\dot\phi}{\Hu}\frac{a^2}{k^2}\tilde{\mu}+\frac{2}{\Hu^2 k^2}\left(\dot\Hu-\Hu\frac{\ddot\phi}{\dot\phi}\right)(a^2\tilde\Upsilon+3\Hu a \tilde\Theta) \nonumber\\
B_0 &= \frac{1}{k\Hu}\left(\kappa a^2 \rho_M - 2 (\Hu^2 - \dot\Hu) \frac{{\tilde\Theta} a}{\Hu}\right)\nonumber\\
C_0 &=2(1- \GTT) -2\frac{\dot{\tilde{\cal G}}_T}{\Hu}\left(1 + 3 \frac{\dot\Hu}{k^2}\right) -6\frac{\GTT}{k^2} \left(2\dot\Hu + \frac{\ddot\Hu}{\Hu}\right) - 
 \frac{3\dot\Hu}{k^2\Hu^2} \kappa a^2 \rho_M\nonumber\\*
 & + \frac{6a\tilde\Theta}{\Hu k^2}\left(4 \dot\Hu- 2 \frac{ \dot{\Hu}^2}{H}+\frac{\ddot\Hu}{\Hu}\right) -\frac{12{\ddot\phi}^2}{k^2{\dot\phi}^2}\left( \GTT-\frac{a\tilde{\Theta}}{\Hu}\right)+\frac{6\dot\Hu a \dot{\tilde\Theta}}{\Hu^2 k^2} - \frac{3 a^2  \tilde{\cal V}\dot\phi}{\Hu k^2} \nonumber\\*
 & \frac{3}{k^2\dot\phi}\left[2\phi^{(3)}\left( \GTT-\frac{a\tilde\Theta}{\Hu}\right)+\ddot\phi\left(2{\dot{\tilde{\cal G}}}_T-\frac{2a\dot{\tilde\Theta}}{\Hu}+4 \GTT\left(\Hu+\frac{\dot\Hu}{\Hu}\right)-8 a\tilde\Theta+\frac{1}{\Hu}\kappa a^2\rho_M\right)\right]\nonumber\\
C_1 &=\frac{2k}{\Hu}(1- \GTT)+\frac{6}{k\Hu}(\Hu^2-\dot\Hu)\left(1-\frac{\tilde\Theta a}{\Hu}\right)\nonumber
\end{align}
\vspace{-5mm}
\begin{align}
D_0 &=1-\GTT-\frac{\dot{\tilde{\cal G}}_T}{\Hu} & D_1 &=\frac{k}{\Hu}(1- \GTT)\nonumber\\
F_0 &=-\frac{2}{k\Hu}(3\Hu^2+a^2{\tilde\Upsilon}) & I_0 &=2\left(1-\frac{\tilde{\Theta} a}{\Hu}\right)\nonumber
\end{align}
\vspace{-2mm}
\begin{align}
J_0 &=\frac{1}{k\Hu}\left[-2k^2(1- \GTT)+3\kappa a^2\rho_M-6\frac{d}{d\eta}(a{\tilde\Theta})+6(\Hu^2+\dot\Hu)-6\frac{\tilde\Theta a}{\Hu}(2\Hu^2-\dot\Hu)\right]\nonumber\\
J_1 &=6\Big(1-\frac{\tilde \Theta a}{\Hu}\Big)\nonumber
\end{align}
\vspace{-3mm}
\begin{align}
K_0 &=-\frac{k}{\Hu}(1- \GTT) & K_1 &=0\nonumber \\
\alpha_0&=M_P\left[\frac{a^2}{k^2}{\tilde\mu}-\frac{2}{\dot\phi}\left(\tilde\Theta a -\Hu\GTT\right)\right]& \alpha_1&=\frac{2 M_P}{k\dot\phi}\left[a^2\tilde\Upsilon+3\Hu\,a\tilde\Theta\right]\nonumber\\
\beta_0&=\frac{M_P}{k \dot\phi^2}\left[-2\ddot\phi\left(a\tilde\Theta-\Hu\GTT\right)-2\GTT\dot\Hu\dot\phi+2a\tilde\Theta\Hu\dot\phi\right]-M_P\frac{\kappa a^2 \rho_M}{k \dot\phi} &\beta_1&=\frac{2M_P}{\dot\phi}\left[\tilde\Theta a -\Hu\GTT\right] \nonumber\\
\gamma_0&=\frac{2M_P}{\dot\phi}\left(\dot{{\tilde{\cal G}}}_T+\Hu\GTT-\Hu\FTT\right)+3M_P\frac{a^2}{k^2}{\tilde{\cal V}} &
\gamma_2&=\frac{6M_P}{\dot\phi}\left(\tilde\Theta a-\Hu \GTT\right)\nonumber
\end{align}
\vspace{-8mm}
\begin{align}
\gamma_1&=\frac{M_P}{k\dot\phi}\left[-6\GTT\left(\dot\Hu+2\Hu^2-2\Hu\frac{\ddot\phi}{\dot\phi}\right)+6\frac{d}{d\eta}\left(a \tilde\Theta -\Hu\GTT\right)+6 a\tilde\Theta\left(3\Hu-2\frac{\ddot\phi}{\dot\phi}\right)-3\kappa a^2 \rho_M\right]& \nonumber
\end{align}
\vspace{-4mm}
\begin{align}
\epsilon_0&=\frac{M_P}{\dot\phi}\left[\dot{\tilde{\cal G}}_T+\Hu\GTT-\Hu\FTT\right] & \epsilon_1&=\epsilon_2=0\label{HD_eps2}
\end{align}
\end{widetext}
$\mu$ and $\cal V$ are derivatives of the zeroth-order field equations with respect to the scalar field (00 and ii components respectively) -  see \cite{DeFelice:2011uq} for the relevant expressions. 

Table \ref{table:HD} collects some `settings' for the Horndeski Lagangian functions that reproduce theories of current interest. The application of the PPF formalism to Horndeski theory immediately brings a large realm of theory space within reach of our parameterization.

\subsection{GR with a Dark Fluid}
\label{subsection:examples_fluid}

Our final example should really be classed as a dark energy model rather than a theory of modified gravity, though arguably the distinction is not important. We present the example of an adiabatic dark fluid characterized by a constant equation of state $\omega_D$ and negligible anisotropic stress. We will use this as an example of how a theory with two additional fields can be recast as a single-field theory (and hence type 1/2/3) in some cases. The example also has relevance to theories that can be usefully written as an effective fluid at the level of the linearized gravitational field equations, e.g. quintessence and its progeny (whilst an effective fluid interpretation is possible for all theories, it is not always useful). 

The zeroth-order modifications to the field equations are simply:
\begin{align}
a^2 X&=\kappa a^2\rho_D & a^2 Y&=\kappa a^2\omega_D\rho_D
\end{align}
The two new fields are the fractional energy density perturbation and the velocity perturbation, $\delta_D$ and $\theta_D$, which appear in the $U$-tensor as:
\begin{align}
U_{\Delta}&=\kappa a^2\rho_D\delta_D & U_{\Theta}&=\kappa a^2\rho_D(1+\omega_D)\theta_D \nonumber\\
U_{P}&=3\kappa a^2\rho_D\omega_D\delta_D & U_{\Sigma}&=0 
\end{align}
The fluid velocity is given by \mbox{$v_D^i=\nabla^i\theta_D$}. For an adiabatic, shear-free fluid the conservation and Euler equations are:
\begin{align}
 \dot{\delta}_D&=-(1+\omega_D)\left(k^2\theta_D+\frac{1}{2}\dot\beta-k^2\epsilon \right) \label{consv} \\
\dot{\theta}_D&=-\Hu(1-3\omega_D)\theta_D+\frac{\omega_D}{1+\omega_D}\delta_D-\Xi \label{fluid_Euler}
\end{align}
These correspond to the two components of the conservation equation \mbox{$\nabla_\mu U^\mu_\nu=0$.}
Eq.(\ref{consv}) can be used to eliminate $\theta_D$ in favour of $\delta_D$ from the gravitational field equations, resulting in a theory with a single new field. After forming the relevant g.i. combination
\begin{align}
\hat{\delta}_D&=\delta_D+\frac{1}{2}(1+\omega)(\beta+k^2\nu) 
\end{align}
the PPF coefficients can be extracted (with all those not stated being zero):
\begin{align}
A_0&=9\Hu_k^2\Omega_D (1+\omega_D) & C_0&=27\Hu_k^2\Omega_D (1+\omega_D)\omega_D \nonumber\\
\alpha^{\hat\delta}_0&=3\Hu_k^2\Omega_D & \beta_1^{\hat\delta}&=-3\Hu_k^2\Omega_D \nonumber\\
\gamma^{\hat\delta}_0&=9\Hu_k^2\Omega_D \omega_D \label{fluid_delta_coeffs}
\end{align}
where $\Omega_D$ is the ratio of the energy density of the dark fluid to the critical density \mbox{$\rho_c=3\Hu^2/(\kappa a^2)$}, as per the usual definition. The superscripts indicate that these coefficients `belong' to the $\hat\delta_D$ perturbations, and should not be confused with spacetime indices.
The e.o.m. is obtained by using eq.(\ref{consv}) to replace $\theta_D$ in eq.(\ref{fluid_Euler}), leading to:
\begin{align}
\ddot{\hat\delta}_D&+\Hu \dot{\delta}_D (1-3\omega_D)+k^2\omega_D\left[\hat\delta_D+3(1+\omega_D)\hat\Phi)\right] \nonumber\\
&+\frac{k^3}{\Hu}\omega_D\hat\Gamma-\frac{k^2}{\Hu}\omega_D\dot{\hat\Phi}=0
\end{align}
Using the PPF coefficients of the modified single-field system, we find that this is our first example of a type 2 theory. That is to say, the coefficients of each term in eq.(\ref{Bianchi_1}) vanish identically.

Of course we can implement an analogous procedure to eliminate $\delta_D$ and treat $\theta_D$ as the single extra field. Note that this cannot be done for the special case \mbox{$\omega_D=0$}, for which we cannot invert eq.(\ref{fluid_Euler}) to find $\delta_D (\theta_D,\dot{\theta}_D)$  (and using eq.(\ref{consv}) instead would lead to integral expressions that do not fit into our parameterization). The relevant gauge-invariant quantity in this case is:
\be
\hat{\theta}_D=\theta_D-\frac{1}{6\Hu}(\beta+k^2\nu)
\ee
Repeating similar steps to before we obtain the PPF coefficients for the new theory, where now $\theta_D$ is the only additional field. Using tildes to indicate that these are not the same as eqs.(\ref{fluid_delta_coeffs}), the results are:
\begin{align}
\tilde{A}_0&=3\Hu_k^2\Omega_D (1+\omega_D)\left[3-\frac{1}{\omega_D}\left(1-\frac{\dot\Hu}{\Hu^2}\right)\right]\nonumber \\
\tilde{B}_0&=-3\Hu_k\Omega_D(1+\omega_D) \nonumber\\
\tilde{C}_0&=9\Hu_k^2\Omega_D (1+\omega_D)\omega_D\left[3-\frac{1}{\omega_D}\left(1-\frac{\dot\Hu}{\Hu^2}\right)\right] \nonumber\\
\tilde{F}_0&=-\frac{(1+\omega_D)}{\omega_D} 3\Hu_k\Omega_D\nonumber\\
\tilde{J}_0&=-9\Hu_k\Omega_D(1+\omega_D)\nonumber\\
\tilde{\alpha}^{k\hat\theta}_0&=3\Hu_k^3\Omega_D (1-3\omega_D)\frac{(1+\omega_D)}{\omega_D} \nonumber\\
\tilde{\alpha}^{k\hat\theta}_1&=3\Hu_k^2\Omega_D \frac{(1+\omega_D)}{\omega_D} &\nonumber\\
\tilde{\beta}^{k\hat\theta}_0&=3\Hu_k^2\Omega_D (1+\omega_D) \nonumber\\
\tilde{\gamma}^{k\hat\theta}_0&=9\Hu_k^3\Omega_D (1-3\omega_D)(1+\omega_D)\nonumber \\
\tilde{\gamma}^{k\hat\theta}_1&=9\Hu_k^2\Omega_D (1+\omega_D) 
\end{align}
This time the 0-component of the Bianchi identity provides the e.o.m.:
 \begin{align}
 \ddot\theta_D&+\Hu\dot\theta_D (1-3\omega_D)+\theta_D \left[\dot\Hu (1-3\omega_D)+k^2\omega_D\right]\nonumber\\
 &+\dot\Xi+\frac{1}{2}(1+\omega_D)(\ddot\beta-2k^2 \dot\epsilon)=0
 \end{align}
The constraints of eqs.(\ref{Bianchi_2_constraints_start}-\ref{Bianchi_2_constraints_end}) are all satisfied. Hence in this formulation the dark fluid becomes a type 1 theory. 

We have at last found a counter-example to the monopoly of type 1 theories. But is this an `artificial' result of moulding a two-field theory into single-field format? We discuss the issue a little further in Appendix \ref{appendix:U_proof}.

\section{The PPF Coefficients}
\label{section:PPF_coeffs}

\noindent 
Different theories leave characteristic signatures in the set of PPF coefficient functions that in principle can be used to distinguish between them. In the first part of this section we discuss the issue of degeneracy between the functions of our framework. In the second part we consider how their scale-dependence should guide the implementation of PPF in numerical calculations.

It is generally argued that the lengthscales relevant to observables such as weak gravitational lensing  and redshift space distortions lie within the `quasistatic' regime. We will discuss the behaviour of the PPF framework in this limit in \textsection\ref{subsection:quasistatic}. In the final part of this section we briefly connect our work to a recent parameterization of screening mechanisms \cite{Braxetal2012}, which are designed to suppress the effects of modified gravity on small scales.

\subsection{Degeneracy}
\label{subsection:degeneracy}
The PPF parameterization employs more free functions than some other similar frameworks in the literature, and it is difficult to predict the degeneracy structure of this set of functions \textit{a priori}. We expect to find that a subset of the PPF functions can be well-constrained, whilst another subset may be more difficult to pin down. We do not expect this to be problematic, as there should still be sufficient information in the well-constrained subset to distinguish between classes of theories.

Furthermore, based on the work of \cite{BattyePearson}, we believe that sets of further relations between the PPF functions exist. These can be derived once the field content of a theory has been specified, and will remove further freedom from the parameterization. This will be in addition to the constraint relations of eqs.(\ref{Bianchi_2_constraints_start}-\ref{Bianchi_2_constraints_end}). We will pursue this point in a future work \cite{Bakerinprep}. There we will also consider in more concrete terms the prospects for discriminating between theories of modified gravity with specific current and future experiments.

The construction of the PPF parameterization necessarily restricts its use to lengthscales at which linear perturbation theory remains valid. However, the advance of errorbars of future experiments towards cosmic variance limits will allow tighter constraints to be obtained even just using the linear window of $k$-space. Furthermore, the advent of new high-redshift probes such as 21cm intensity mapping will give access to early times when a wider range of scales fell within the linear regime.

To make use of new data probing smaller scales one would need to use a prescription for nonlinear corrections, such as \cite{Samsing_Linder}. Applying such a prescription requires calibration from N-body simulations of specific modified gravity models, and this undermines the model-independent approach that we are pursuing. It would also introduce further parameters to be marginalized over in a data analysis of the PPF formalism; given the possible degeneracies already present, we choose not to use nonlinear-scale data (\mbox{$k\gtrsim 0.2 \,\mathrm{h}^{-1}$Mpc} at $z=0$) at this time.
\begin{center}
\begin{table*}
\begin{tabular}{| c | c | c | c | c | c |}\hline
 & \bf{1} & \bf{2} & \bf{3} & \bf{4} &  \\ \hline
\;\bf{Theory}\; & \;\bf{$(A_0+3\Hu_k B_0)\,k^2\hat\Phi$} \;&\; \bf{$(F_0+3\Hu_k I_0)\,k^2\hat\Gamma$} \;&\; \bf{$(\alpha_0+3\Hu_k \beta_0)\,k^2\hat\chi$} \;& \;\bf{$(\alpha_1+3\Hu_k \beta_1)\,k\dot{\hat\chi}$} \;&\; \bf{Slip}\; \\ \hline 
Scalar-Tensor & $k^0, \,k^2$ & $k^0$ &  $k^0,\,k^2$ & $k^0$ & $k^0$ \\ \hline
Einstein-Aether & $k^2$ & $k^0,\,k^2$ &  $k^2$ & $k^2$ & $k^0$ \\ \hline
DGP & $k^0, \,k^2$ & $k^0$ &  $k^0$ & $k^0$ & $k^0$\\ \hline
EBI & $k^0$ & - &  $k^0$ & $-$ & $k^0$\\ \hline
Ho\u{r}ava-Lifschitz & $k^2, \,k^4,\,k^6$ & $k^0, k^2, \,k^4,\,k^6$ &  $k^2, \,k^4,\,k^6$ & $k^2, \,k^4,\,k^6$ & $k^0,\,k^2,\,k^4$\\ \hline
Horndeski & $k^0, \,k^2$ & $k^0$ &  $k^0,\,k^2$ & $k^0$& $k^0$ \\ \hline
Fluid - $\hat{\delta}_D$ form & $k^0$ & - &  $k^0$ & $k^{-2}$ & $-$ \\ \hline
Fluid - $\hat{\theta}_D$ form & $k^0$ & $k^0$ &  $k^0$ & $k^0$ & $-$ \\ \hline
\end{tabular}
\label{table:k_dep}
\caption{The scale-dependence of the terms appearing in the modified Poisson equation (\ref{PPF_Poisson}) for the example theories of \textsection\ref{section:examples}. Recall that $\hat\Gamma \propto k^{-1}$, see eq.(\ref{Gamma_def}). Note that in some cases the scalar perturbation $\chi$ includes a factor of $k$ so that it is dimensionless, eg. $\chi=k{\xi^0}$ in Ho\u{r}ava-Lifshitz gravity, $\chi=k\theta$ for a fluid velocity potential. EBI and DGP gravity will contain multiple copies of terms 3 and 4 (eg. \mbox{$\hat{\chi}_1=\hat{\delta}_W$},  \mbox{$\hat{\chi}_2=k\hat{\theta}_W$} for DGP), but both terms have the same scale-dependence.}
\end{table*}
\end{center}

\subsection{Signatures and Scale-Dependence}
\label{subsection:scale_dep}
%
Naively, one might expect the well-constrained subset of PPF functions to be those which feature directly in the calculation of observable quantities. One rarely needs to use the full set of Einstein equations to calculate observables; usually 
%
%
only the Poisson equation, the slip relation (\mbox{$\Phi-\Psi$}), and the evolution equations for $\delta_M$ and $\theta_M$ are required. The modified Fourier-space Poisson equation in the PPF parameterization is (taking eq.(\ref{FE1})+$3\Hu\,\times\,$eq.(\ref{FE2})):
\begin{align}
-2k^2\hat\Phi&=\kappa a^2\rho_M\Delta_M \nonumber \\
&+{(A_0+3\Hu_k B_0)\,k^2\hat\Phi}&+&{(F_0+3\Hu_k I_0)\,k^2\hat\Gamma} \nonumber\\
&+ {(\alpha_0+3\Hu_k \beta_0)\,k^2\hat\chi}&+&{(\alpha_1+3\Hu_k \beta_1)\,k\,\dot{\hat\chi}}
\label{PPF_Poisson}
\end{align}
where $\Delta_M=\delta_M+3\Hu (1+\omega_M)\theta_M$ is a gauge-invariant density perturbation.

Table \ref{table:k_dep} indicates the scale-dependence of the non-standard terms of eq.(\ref{PPF_Poisson}) in each of the theories of \textsection\ref{section:examples}. It is clear that modifications to the Poisson equation have the form of a power series in even powers of the Fourier wavenumber $k$. It is worth stressing that this form is exact, and not a Taylor series expansion of a more complicated function; an even power series in $k$ is the only possibility. In most cases only the scale-free and quadratic terms are present, in agreement with the findings of \cite{Wagoner:2008vg} (apart from DGP, as discussed in \textsection\ref{subsection:examples_dgp}). This is no surprise - we have focused on theories that are second order in time derivatives, and Lorentz invariance then implies that they must be second order in spatial derivatives too. Odd powers of $k$ are forbidden by parity. 

Ho\u{r}ava-Liftschitz gravity is a notable exception - its explicit breaking of the symmetry between time and space coordinates allows higher-order spatial derivatives to enter the PPF coefficients. 
%
%
%
Interestingly, all the theories except Ho\u{r}ava-Lifshitz gravity are found to have scale-independent slip relations (for brevity we have not decomposed the slip relation into its individual terms in Table \ref{table:k_dep}). The coefficients $K_1$ and $\epsilon_2$ are zero in all cases. 

Based on these observations there are three ways to implement the PPF coefficients in an Einstein-Boltzmann solver code such as the CAMB package \cite{camb}:
\vspace{1mm}
\begin{enumerate}[i)]
\item We can construct a sensible parameterization of the PPF coefficients by splitting them into purely time-dependent functions multiplying a spatial dependence. To be fully general (at least in terms of the theories treated in this paper) we should expand each of the PPF functions appearing in the Poisson equation as:\newline
\begin{align}
\label{ansatz}
\quad\quad A_0=\sum_{n=0}^4\tilde{A}_0(z)\,k^{2(n-1)}
\end{align}
and similarly for $B_0,\,F_0$ etc. One must choose a motivated ansatz for the time-dependent function $\tilde{A}_0 (z)$, such as a Taylor series in $\Omega_\Lambda$ or $(\Hu_0/\Hu)^{-1}$. In reality, a form as general as eq.(\ref{ansatz}) is probably unnecessary for all of the coefficients. 

\item Alternatively one may employ a Principal Component Analysis (PCA), as advocated in \cite{Zhao:2009eq,Pogosian:2010hi,Hojjati:2011vp,Zhao:2012wx}. The PCA approach expands the functions to be constrained in a basis of orthogonal eigenmodes in $\{k,z\}$-space, using the eigenbasis that a given data set is most sensitive to. Informally speaking, PCA reveals `what the data knows', rather than `what we would like to know'. 
\item A compromise between the two methods above is possible: expand the free functions as in eq.(\ref{ansatz}), then perform a one-dimensional PCA for the unknown functions of redshift.
\end{enumerate}

Method i) has the advantage of simplicity, and it utilizes the common features of modified gravity theories to jump directly to the most relevant regions of theory space. However, by choosing a form for functions such as $\tilde{A}_0 (z)$ we are imposing our preconceptions on the behaviour of modified gravity, namely that it must emerge at late times (z $\lesssim$ 1) in order to reproduce the effects of an apparent dark energy. 

Method ii) maintains a greater degree of agnosticism, avoiding the use of semi-arbitrary functional forms. However, it effectively throws away much of our prior knowledge about the structure of physically-reasonable theories. 

%
We are most likely to pursue method iii), which represents a compromise between constraining known regions of the modified gravity landscape and exploring new unknown territory. For the sake of argument, let us suppose we use $n=1,2$ for terms 1 and 3 in Table \ref{table:k_dep} and $n=1$ for terms 2 and 4. The Poisson equation is then described by six functions of redshift. Since $K_1=\epsilon_2=0$ for all theories (a result of the additional constraint equations we alluded to in \textsection\ref{subsection:degeneracy}), the slip function contains five functions of redshift (since all terms in it are scale-independent). This totals eleven functions of redshift only to constrain.  The authors of \cite{Hojjati:2011vp} find that combining results from the Planck satellite with an LSST-like experiment yields 155 data points, which gives us fourteen data points per function. This should be sufficient to pin down a few eigenvalues for each of the eleven functions.

\subsection{The Quasistatic Limit}
\label{subsection:quasistatic}

On sufficiently subhorizon scales ($\Hu_k \ll 1$) -- but above the nonlinearity scale -- the time derivatives of metric perturbations are small relative to their spatial derivatives, and can be neglected (at least for the purposes of N-body simulations). This is known as the quasistatic regime \cite{Hu:2007fw}. It is believed that time derivatives of the additional fields in modified gravity theories can be similarly neglected in this regime: many scalar-field type models exhibit damped oscillatory solutions below a special lengthscale, which rapidly decay into insignificance. Only the averaged, stationary spatial profile then remains of relevance. 

Let us consider type 1 theories, as these are the most prevalent. In the quasistatic limit eq.(\ref{Bianchi_1}), the e.o.m. for the new scalar d.o.f., reduces to a relation between the perturbations of the extra field and the potential $\hat\Phi$:
 \be
 \label{QS_chi_rel}
\left[\dot{\alpha}_0+k\beta_0\right]\hat\chi+\left[\dot{A}_0+k B_0\right]\hat\Phi=0
\ee
Time derivatives of the PPF functions have been maintained because the quasistatic approximation only allows us to neglect time derivatives of \textit{perturbations}. To obtain the quasistatic limit of the Poisson equation we use eq.(\ref{QS_chi_rel}) to replace $\hat\chi$ in eq.(\ref{PPF_Poisson}), and drop the $\dot{\hat\chi}$ and $\hat\Gamma$ terms (recall that \mbox{$\hat\Gamma=1/k \,(\dot{\hat\Phi}+\Hu\hat\Psi)$}, where $\dot{\hat\Phi}$ will be small and $\hat\Psi$ is suppressed by $\Hu_k$). We apply a similar treatment to the slip relation, neglecting the anisotropic stress of matter. The results can be written in the form:
\begin{align}
-2k^2\hat\Phi&=\kappa a^2 \mu (a,k)\,\rho_M\Delta_M \\
\frac{\hat\Phi}{\hat\Psi}&=\gamma (a,k)
\end{align}
where
\begin{align}
\mu (a,k)&=\left[1+\frac{A_0}{2}-\frac{\alpha_0}{2}\left(\frac{\dot{A}_0+k B_0}{\dot{\alpha}_0+k\beta_0}\right)\right]^{-1}\\*
\gamma (a,k)&=\left[1-D_0+\epsilon_0\left(\frac{\dot{A}_0+k B_0}{\dot{\alpha}_0+k\beta_0}\right)\right]^{-1}
\end{align}
We see that on quasistatic scales modifications to the Poisson and slip relations can be reparameterized in terms of two time- and scale-dependent functions, $\mu (a,k)$ and $\gamma (a,k)$. This form of parameterization has been explored in a large number of investigations \cite{Hu:2007fw,Bertschinger:2008bb,Linder:2008uv, Pogosian:2010hi,Daniel:2010gt,Bean:2010kk, Dossett:2011vu,Thomas:2011tc}, using a number of different ansatzes for the scale-dependence of $\{\mu,\gamma\}$. From the discussion of the previous subsection we see that one does not really have the freedom to choose just any ansatz here; with a few exceptions, the physically-relevant choice is that of scale-free terms plus $k^2$ terms, such as used in \cite{Bertschinger:2008bb}. 

Deviations of $\{\mu,\,\gamma\}$ from their GR values of $\{1,\,1\}$ are degenerate between modifications to the metric sector and new d.o.f. sector, indicated by the presence of Greek and Latin PPF coefficients in both. The advantage of maintaining the unnapproximated PPF Poisson equation (eq.(\ref{PPF_Poisson})) is that these coefficients are kept distinct, which is necessary if we hope to detect the signatures of particular theories discussed in the previous subsection. Keeping terms such as $\dot{\hat\chi}$ in the equations may be unfeasible or unnecessary for N-body simulations, but it should not pose a problem for Einstein-Boltzmann codes. 
Furthermore, lifting the quasistatic approximation means that the PPF parameterization is valid all the way up to horizon scales, which are relevant for large-scale CMB modes contributing to the late-time Integrated Sachs-Wolfe effect and the lensing-ISW cross-correlation \cite{DiValentino:2012tp, Hu_Liguori_2012}.

\subsection{Connection to Screening Parameterizations}

Brax and collaborators have introduced a new parameterization \cite{Braxetal2012, Brax:2012wq} that is optimized for screening mechanisms, such as the chameleon \cite{2004PhRvD..69d4026K,2004PhRvL..93q1104K}, dilaton \cite{Brax_dilaton} 
and symmetron mechanisms \cite{2011PhRvD..84j3521H,2010PhRvL.104w1301H}.  In this parameterization a theory is described by the time-evolving mass of the scalar field, $m(a)$, and its coupling to matter in the Einstein frame, $\beta (a)$, where the Jordan and Einstein metrics are related by:
\be
g^\mathrm{J}_{\mu\nu}&=A^2 (\phi)\, g_{\mu\nu}^\mathrm{E} 
\ee
and the coupling function is:
\be
\beta (a)=\frac{d\,\mathrm{ln} A(\phi)}{d\phi}
\ee
The parameterization is constructed explicitly in the Einstein frame; this means that when mapped to our (Jordan-frame) parameterization it mixes PPF coefficients from the metric and extra-field sectors. In the quasistatic regime the mapping is:
\begin{align}
\left[1+\frac{A_0}{2}-\frac{\alpha_0}{2}\left(\frac{\dot{A}_0+k B_0}{\dot{\alpha}_0+k\beta_0}\right)\right]^{-1}&=1-\frac{2\beta^2}{1+m^2\frac{a^2}{k^2}}\\
\left[1-D_0+\epsilon_0\left(\frac{\dot{A}_0+k B_0}{\dot{\alpha}_0+k\beta_0}\right)\right]^{-1}&=1+\frac{2\beta^2}{1+m^2\frac{a^2}{k^2}}
\end{align}
Adding the two expressions above gives an additional constraint relation between the PPF coefficients in the quasistatic regime.

If we lift the quasistatic approximation then in principle there is sufficient information to express the twenty-two PPF functions in terms of $\beta (a)$ and $m(a)$. However, the expressions are not particularly elegant and we will not present them here on grounds of relevance: recent work \cite{WangHui2012} indicates that the scalar fields involved in the chameleon, dilaton and symmetron screening mechanisms must have a Compton wavelength $\lesssim 1$ Mpc in order to satisfy Solar System and Galactic constraints. Deviations from GR are then only expected in the nonlinear regime, to which PPF does not apply.

\section{Conclusions}
\label{section:conclusions}

We have presented a new framework, the Parameterized Post-Friedmann formalism (PPF), for conducting model-independent tests of modified gravity. The construction of the framework does not rely on ad-hoc modifications to GR, but is built by considering the limited number of ways that the linearized Einstein equations can be extended whilst maintaining the properties of a physically-relevant gravitational theory. An exact, analytic mapping exists between a theory of modified gravity that obeys the assumptions of Table \ref{table:assumptions} and the set of PPF coefficient functions. 

There are four key reasons we believe that the PPF formalism represents a significant step forward in parameterized treatments of modified gravity:
\begin{enumerate}
\item New degrees of freedom, a common feature of many models, are explicitly parameterized for. Until recently \cite{Braxetal2012} this has not generally been the case. This is not a minor issue; new degrees of freedom are at the heart of the phenomenology that might render a theory of modified gravity distinguishable from $\Lambda$CDM.
\item The parameterization is not restricted to models based solely on scalar fields. As experimental precision increases, the continued success of $\Lambda$CDM has forced model-builders to look to richer theories that additionally involve vector and tensor fields. PPF is able to handle the spin-0 perturbations of these theories as easily as canonical scalar fields.
\item Once cast into the format of PPF, a theory of modified gravity may leave a distinctive `calling card' in the set of PPF functions. Such signatures may allow some theories to be ruled out, or at least indicate the general characteristics of theories that are consistent with the data, eg. one expects pure scalar field models to share similarities, whilst Lorentz-violating theories may share similar traits. It seems reasonable to suppose that higher-dimensional models may have their own unique features.
\item The parameterization is valid from horizon scales down to the scales at which nonlinearities become important. This spans a wide range of observables targeted by the next generation of experiments \cite{DES,LSST,EIC,SKA}, including weak lensing (of both galaxies and the CMB), redshift space distortions, peculiar velocity surveys, the ISW effect and associated cross-correlations, and possibly galaxy clusters \cite{Thomas:2011tc}. 
\end{enumerate}

The PPF framework can be applied in two modes depending on the user's interests. It can be used as a tool for tackling observations, a gateway that allows data constraints to be applied to a multitude of `known' theories simultaneously. By `known' theories we mean existing models for which we have the field equations in hand. Alternatively, one can use the framework as an exploratory tool for model-building. As mentioned in 3) above, PPF may be able to indicate the physical features that lead to tension with the data, thereby guiding theorists to the regions of theory space that are likely to prove most fruitful. We note that for theories involving more than two new fields the exploratory mode cannot be used, but the observational mode remains fully functional.

The cost of the increased capabilities of the PPF framework is the introduction of more `free' functions than other parameterizations (though we have seen in this paper that the structure of these functions is most definitely not arbitrary). In \textsection\ref{subsection:formalism_constraints} we presented a number of constraints that can be used to reduce this freedom, and we believe that there exist further relations not presented here \cite{Bakerinprep}. The situation is likely to be analogous to that of the Parameterized Post-Newtonian framework, where a set of ten new potentials and ten new parameters are necessary to capture nearly all possible distortions of the GR metric: it turns out that many of these new parameters are needed in only a handful of special cases, and are zero the rest of the time. 

We have been ambitious. It is possible that the current and next-generation datasets may not be able to tease apart the degeneracies between contributions to the modified field equations to the extent required here. In this situation the PPF framework still functions as a null test for non-GR behaviour, even if it cannot reach its full potential as a discriminator of modified gravity theories. Then the most pertinent question to ask is: how much better do our experiments need to be to realize this discriminatory power? 
Tackling this question requires implementation of the PPF framework in a numerical code for computing cosmological observables. The results of this implementation -- and a possible answer to our question -- will be presented in a future work \cite{Bakerinprep}.

\vspace{-10pt}
\section*{Acknowledgements}
\vspace{-10pt}
We are grateful for helpful discussions with Joanna Dunkley, Kazuya Koyama, Tony Padilla, Jonathan Pearson, Jim Peebles, Levon Pogosian, Thomas Sotiriou, Clifford Will, Gong-Bo Zhao and Tom Zlosnik. 

TB is supported by the STFC.  PGF acknowledges support from the STFC, the Beecroft Institute for Particle Astrophysics and Cosmology, and the Oxford Martin School. CS is supported by a Royal Society University Research Fellowship.

\appendix

\section{Transformation of $f(R)$ gravity into a Scalar-Tensor Theory}
\label{appendix:fR_transf}

\noindent Consider the action:
\begin{equation}
\label{fR_equiv_action}
S=\int\,d^4x\, \sqrt{-g}\left[f(\zeta)+f^{\prime}(\zeta)(R-\zeta)\right]+S_M (\psi^a, g_{\mu\nu})
\end{equation}
where primes denote differentiation with respect to the new scalar $\zeta$. 
Variation of this action gives the field equation for $\zeta$ as:
\begin{equation}
f^{\prime\prime}(\zeta)(R-\zeta)=0
\end{equation}
Hence $\zeta=R$ for all \mbox{$f^{\prime\prime}(\zeta)\neq 0$}. Substituting $R=\zeta$ into eq.(\ref{fR_equiv_action}) recovers the standard $f(R)$ action. Therefore eq.(\ref{fR_equiv_action}) is an equivalent action for $f(R)$ gravity, with the special case of \mbox{$f^{\prime\prime}(\zeta)= 0$} corresponding to the Einstein-Hilbert action.

To recast the equivalent action as a scalar-tensor theory we make the definitions:
\begin{align}
\phi &=\frac{d\,f(R)}{dR}=f_R & V(\phi)=\frac{1}{2}\left[\zeta \phi-f(\zeta)\right]
\end{align}
Then eq.(\ref{fR_equiv_action}) becomes:
\begin{equation}
\label{fR_st_action}
S=\int\,d^4x\, \sqrt{-g}\left[\phi R-2 V(\phi)\right]+S_M (\psi^a, g_{\mu\nu}) 
\end{equation}
which has the form discussed in \textsection\ref{subsection:examples_st}, with $\omega (\phi)=0$ in this case.

Note that because the metric has not been transformed its couplings to the scalar field and matter are unaffected.

\section{Is the Bianchi 0-Component Always the E.O.M.? A Plausibility Argument}
\label{appendix:U_proof}

\noindent In \textsection\ref{subsection:formalism_constraints} we introduced a classification scheme for single-field theories, according to whether the spatial or temporal part of the perturbed conservation law $\delta (\nabla_\mu U^\mu_\nu)=0$ reduced to a trivial relation (types 1 and 2 respectively), or neither component did (type 3). Of the seven examples considered in \textsection\ref{section:examples} we found that four were type 1 (Scalar-tensor, Einstein-Aether, Ho\u{r}ava-Lifshitz and Horndeski). EBI and DGP (the effective 4D theory) were are not subject to this classification, having more than two new fields. The adiabatic, shear-free dark fluid could be recast as either type 1 or type 2.

It seems, then, that type 1 theories dominate. Why is this? We have yet to find an example of a type 3 theory, and it could be argued that our example of a type 2 theory is the result of an artificial re-writing. We present here a plausibility argument why \textit{all} single-field theories may be type 1, but do not make any claims that it is a fully rigorous or watertight proof.

Consider the following action:
\begin{equation}
S=\int\left[\frac{R}{2\kappa}+{\cal L}_M+\frac{1}{\kappa}{\cal L}_U\right]\rtg \;d^4x
\label{action}
\end{equation}
where ${\cal L}_U$ is the Lagrangian that gives rise to the $U$-tensor that appears in the gravitational field equations, eq.(\ref{schematic_U}). We assume that any gravitational action can be written in the form of eq.(\ref{action}) by suitable choice of ${\cal L}_U$, e.g. for $f(R)$ gravity we would have ${\cal L}_U=\frac{1}{2}[f(R)-R]$. The variation of the last term in eq.(\ref{action}) is:
\begin{equation}
\label{linear_action}
\delta S_U=\frac{1}{2\kappa}\int U^{\mu\nu}\delta g_{\mu\nu} \rtg \,d^4x
\end{equation}
where
\begin{equation}
\label{U_def}
U^{\mu\nu}=-\frac{2}{\rtg}\frac{\delta(\rtg\; {\cal L}_U)}{\delta g_{\mu\nu}}
\end{equation}

Now consider an infinitesimal co-ordinate transformation, $x^{\mu}\rightarrow x^{\mu}+\xi^{\mu}(x)$. The corresponding change in the metric is given by $\delta g_{\mu\nu}=-\nabla_{\mu}\xi_{\nu}-\nabla_{\nu}\xi_{\mu}$. Substituting this expression into eq.(\ref{linear_action}) and utilizing the symmetry between indices $\mu$ and $\nu$: 
\begin{eqnarray}
\label{pre_Bianchi_action}
\delta S_U&=&-\int \left(\nabla_{\mu}\xi_{\nu}\right)\,U^{\mu\nu}\,\rtg \,d^4x\\
&=&-\int \nabla_{\mu}\left(\xi_{\nu}\,U^{\mu\nu}\right) \rtg \,d^4x \nonumber\\
&&+\int \xi_{\nu}\left(\nabla_{\mu}U^{\mu\nu}\right) \rtg \,d^4x
\label{Bianchi_action}
\end{eqnarray}
The first term of eq.(\ref{Bianchi_action}) can be converted into a boundary integral via the usual divergence theorem. This boundary term must equal zero, because it is an assumption of the variational procedure that the small variations $\xi_{\nu}$ vanish at spatial infinity.

Demanding stationarity of the action, the second term of eq.(\ref{Bianchi_action}) yields the expression:
\begin{equation}
\xi^{\nu}\nabla_\mu U^\mu_\nu =0
\end{equation}
which is easily recognized as the conservation law for $U^{\mu\nu}$ contracted with the gauge transformation vector. We obtain the two components of the conservation law for $U_{\mu\nu}$ by considering two separate coordinate transformations, a temporal one ($x^{0}\rightarrow x^{0}+\xi^{0}$) and a spatial one ($x^{i}\rightarrow x^{i}+\xi^{i}$). The first corresponds to translation along a worldline, and the second to translation of spatial coordinates across a fixed-time hypersurface. A vector proportional to $\xi^0$ is not a Killing vector of the FRW spacetime, so it is not surprising that a coordinate translation along $\xi^0$ results in an \textit{evolution} equation for the new degrees of freedom. In contrast, spatial vectors proportional to $\xi^i$ are Killing vectors of the FRW metric - so no dynamics can be obtained from a translation along $\xi^i$. Hence the terms of the $\nu=i$ equation must vanish identically.


\bibliographystyle{apsrev}


\end{document}